\documentclass[aps,prd,twocolumn,nofootinbib,superscriptaddress,preprintnumbers,floatfix]{revtex4-1}
\usepackage{amsfonts,amsmath,amssymb,amsthm,braket,nicefrac}
\usepackage{booktabs,array}
\usepackage[utf8]{inputenc}
\usepackage{float,graphicx,hhline}
\usepackage{appendix}
\usepackage[dvipsnames]{xcolor}
\usepackage{mathtools}
\usepackage{mathrsfs}

\usepackage{hyperref}
\hypersetup{
    pdftitle={},
    colorlinks=true,     
    linkcolor=blue,      
    citecolor=blue,      
    filecolor=blue,      
    urlcolor=blue        
}
\usepackage{cleveref}

\newcommand{\xx}{\mathbf{x}}
\newcommand{\yy}{\mathbf{y}}

\newcommand{\Tr}{\mathrm{Tr}\,}

\newcommand{\Dgm}{\mathrm{Dgm}}
\newcommand{\EE}{\mathbf{E}}
\newcommand{\zz}{\mathbb{Z}}
\newcommand{\rr}{\mathbb{R}}
\newcommand{\nn}{\mathbb{N}}
\newcommand{\BB}{\mathbf{B}}
\newcommand{\Pp}{\mathcal{P}}
\newcommand{\Bb}{\mathcal{B}}
\newcommand{\Cc}{\mathcal{C}}
\newcommand{\Dd}{\mathcal{D}}

\newcommand{\getHeidelbergAffiliation}{\affiliation{Institut für Theoretische Physik,	Universität Heidelberg, Philosophenweg 16, D-69120 Heidelberg, Germany}}
\newcommand{\getEMMIAffiliation}{\affiliation{ExtreMe Matter Institute EMMI, GSI, Planckstr. 1, D-64291 Darmstadt, Germany}}
\newcommand{\getMITAffiliation}{\affiliation{Center for Theoretical Physics, Massachusetts Institute of Technology, Cambridge, MA 02139, USA}}
\newcommand{\getIAIFIAffiliation}{\affiliation{The NSF AI Institute for Artificial Intelligence and Fundamental Interactions}}

\begin{document}

\title{Confinement in non-Abelian lattice gauge theory via persistent homology}

\author{Daniel Spitz}
\email{spitz@thphys.uni-heidelberg.de}
\getHeidelbergAffiliation

\author{Julian M. Urban}
\getMITAffiliation
\getIAIFIAffiliation

\author{Jan M. Pawlowski}
\getHeidelbergAffiliation
\getEMMIAffiliation

\preprint{MIT-CTP/5512}

\begin{abstract}
We investigate the structure of confining and deconfining phases in SU(2) lattice gauge theory via persistent homology, which gives us access to the topology of a hierarchy of combinatorial objects constructed from given data. Specifically, we use filtrations by traced Polyakov loops, topological densities, holonomy Lie algebra fields, as well as electric and magnetic fields. This allows for a comprehensive picture of confinement. In particular, topological densities form spatial lumps which show signatures of the classical probability distribution of instanton-dyons. Signatures of well-separated dyons located at random positions are encoded in holonomy Lie algebra fields, following the semi-classical temperature dependence of the instanton appearance probability. Debye screening discriminating between electric and magnetic fields is visible in persistent homology and pronounced at large gauge coupling. All employed constructions are gauge-invariant without a priori assumptions on the configurations under study. This work showcases the versatility of persistent homology for statistical and quantum physics studies, barely explored to date.
\end{abstract}

\maketitle

\section{Introduction}

Many non-perturbative phenomena are driven by topological configurations or rather their density, or are accompanied by qualitative changes in the latter. This makes topology changes an ideal probe for investigating the mechanisms and signatures of these phenomena, ranging from (topologically driven) phase transitions to non-perturbative scalings as seen in the presence of topological tunneling effects in the anharmonic oscillator. A prominent and important example is quantum chromodynamics (QCD), whose confinement-deconfinement phase transition is accompanied by a rapid change in the topological density, and anomalous chiral symmetry breaking is closely related to instantons---stable, classical (minimal-action) configurations. Topological confinement mechanisms have been suggested, based on topological defects in QCD such as vortices and monopoles. While not being finite-action configurations in QCD, monopoles and vortices naturally emerge as constituents of instantons. For the discussion of finite temperature instantons (calorons), see, e.g.,~\cite{Harrington:1978ve, kraan1998periodic, kraan1998monopole, lee19982}. More generically, instanton constituents of different topological types emerge on general compact manifolds~\cite{Ford:2000zt, Ford:2002pa, Ford:2003vi, Ford:2005sq}, which may be seen as a simulation with a given topological density. 

At finite temperature, instanton-dyons interact with holonomies (Polyakov loops)~\cite{kraan1998periodic, kraan1998monopole, lee19982}, and have been identified in pure lattice gauge theory and lattice QCD~\cite{perez1999calorons, ilgenfritz2002topological, bornyakov2013topology, bornyakov2016dyons, bornyakov2017dyons}. Ensembles of instanton-dyons can often readily explain confinement~\cite{gerhold20072, diakonov2007confining, bruckmann2012confining, larsen2015interacting, liu2015confining, lopez2018confinement}, even for theories with exceptional gauge groups exhibiting a trivial center such as G(2)~\cite{pepe2007exceptional, diakonov2009topology, diakonov2011confinement}, yet consistently giving rise to center symmetry breaking for theories with non-trivial center. Local correlations exist between topological hotspots and values of the Polyakov loop trace~\cite{larsen2022correlating}. Moreover, instantons and instanton-dyons can be linked to spontaneous chiral symmetry breaking in QCD if sufficiently dense~\cite{liu2015light, larsen2016instanton, demartini2021chiral}. 

Evidently, phase transitions go hand in hand with a topology change in the vacuum manifold, and more generally that of equipotential hypersurfaces~\cite{franzosi2004theorem, franzosi2007topology1, franzosi2007topology2, Gori:2018etr}. Persistent homology has been developed to detect topological structures in finite, noisy data~\cite{carlsson2009topology}, accompanied by profound mathematical works on their stability~\cite{cohen2007stability, cohen2010lipschitz, bauer2013induced} and benign statistical behavior~\cite{hiraoka2018limit, spitz2020self}. From the data, a hierarchy of combinatorial objects is constructed, whose topology can be algorithmically computed using homology. The hierarchical information provides a means to discriminate topological structures according to their dominance (persistence). In light of topology changes of equipotential hypersurfaces in the realm of phase transitions, emergent topological structures have been searched for numerically via persistent homology in configuration space~\cite{donato2016persistent}. Since the scalability of this approach to larger systems is unclear, studies of a variety of condensed matter and spin systems have since been focusing on persistent homology as sensitive observables in their own right~\cite{speidel2018topological,santos2019topological, olsthoorn2020finding,tran2021topological, cole2021quantitative, tirelli2021learning, sale2022quantitative,he2022persistent}. Using a specifically tailored filtration constructed from plaquettes, center vortices have been probed in SU(2) lattice gauge theory~\cite{Sale:2022qfn}. Effective QCD models have also been investigated~\cite{Hirakida:2018bkf, Kashiwa:2021ctc}. Further persistent homology applications in physics include non-thermal fixed points~\cite{spitz2021finding}, quantum entanglement~\cite{olsthoorn2021persistent}, physical chemistry and amorphous materials~\cite{duponchel2018exploring, nakamura2015persistent}, the cosmic web~\cite{sousbie2011persistent, weygaert2011alpha, pranav2017topology}, and non-Gaussianities in cosmic microwave background fluctuations~\cite{pranav2019unexpected, biagetti2021persistence}. Quantum algorithms for the computation of persistent homology have also been developed~\cite{lloyd2016quantum}. Furthermore, topological neural network layers can be constructed using persistent homology~\cite{adams2017persistence, pun2022persistent}.

Typically, the study of classical field configurations in lattice gauge theories requires the application of cooling/smoothing techniques, as well as sophisticated gauge-fixing and \mbox{-p}rojection procedures~\cite{rothe2012lattice}. To identify instanton-dyons on the lattice, overlap fermions may be used as probes~\cite{larsen2020towards}. In the present work, we set out to study finite-temperature pure SU(2) gauge theory on a four-dimensional Euclidean lattice via observables constructed from persistent homology. All of the investigated order parameters are gauge-invariant. In addition, our approach is not biased towards particular classical field configurations. The Hybrid/Hamiltonian Monte Carlo (HMC) algorithm~\cite{DUANE1987216} is employed to generate samples. To relate structures occurring in the full quantum theory to classical field configurations, a comparison with cooled configurations is provided. Using persistent homology, we investigate sub- and superlevel set cubical complex filtrations of different gauge-invariant local observables such as Polyakov loop traces, topological densities, Polyakov loop algebra element norms, as well as electric and magnetic field strengths. This allows for diverse insights into the (non-)local structures occurring at different couplings: topological densities form spatial lumps instead of string-like structures, approximately invariant under cooling in the confined phase and thus \mbox{(near-)classical}. Probabilistic predictions for the appearance of instantons and instanton-dyons are met by persistent homology quantifiers. Debye screening discriminates between spatial structures of electric and magnetic fields. The identification of the approximate critical coupling is facilitated by the simultaneous appearance of qualitative changes across persistent homology observables.

This paper is organized as follows. We review relevant aspects of lattice gauge theory calculations as well as the Polyakov loop trace as the common confinement order parameter in \Cref{SecLatticeIntro}. \Cref{SecPolyakPersHom} deals with the persistent homology of different Polyakov loop descriptors, and begins with a description of (filtered) cubical complexes and persistent homology. Results for the traced Polyakov loop filtration, the Polyakov loop topological density filtration, and the so-called angle-difference filtration of Polyakov loop algebra element norms are discussed. In \Cref{SecElectrMagnPersHom}, we investigate filtrations of traced electric and magnetic field strengths, as well as the topological density. Finally, we conclude and issue an outlook in \Cref{SecConclusions}.

\section{Order parameters from lattice calculations}\label{SecLatticeIntro}

We provide a brief description of our lattice setup in \Cref{SecLatticePrerequisites}. In \Cref{SecPolyakovIntro}, we discuss standard order parameters for the confinement-deconfinement phase transition, based on the Polyakov loop.

\subsection{Prerequisites}\label{SecLatticePrerequisites}

We consider a four-dimensional Euclidean lattice with $N_\sigma^3 \times N_\tau$ sites and periodic boundary conditions in all directions. We denote the set of all lattice sites by $\Lambda$ and the spatial $N_\sigma^3$ lattice given by the coordinates $(n_x,n_y,n_z,0)$ by $\Lambda_\sigma$. Throughout this work, we fix $N_\sigma = 32$ and $N_\tau = 8$, but aim to investigate different and larger lattice geometries as well as conduct a detailed analysis of the $N_\tau$-dependence in the future.

The gauge degrees of freedom are described in terms of SU(2)-valued links denoted as $U_\mu(x)$. Under a local gauge transformation $V(x)$, links transform as $U_\mu(x) \mapsto V(x) U_\mu(x) V^\dagger(x+\hat{\mu})$. With $\beta$ the inverse coupling squared and the plaquettes $U_{\mu\nu}(x) = U_\mu(x) U_\nu(x+\hat{\mu}) U^\dagger_\mu(x+\hat{\nu}) U^\dagger_\nu(x)$, the Wilson gauge action reads
\begin{equation}\label{EqWilsonGaugeAction}
    S[U] = \frac{\beta}{2}\sum_{x\in\Lambda} \sum_{\mu < \nu}\Tr [1 - U_{\mu\nu}(x)]
\,.
\end{equation}
From lattice renormalization group arguments, close to the critical point a linear relation exists between $\beta$ and the temperature~\cite{velytsky2008finite}. We evaluate observables at 16 evenly spaced points ranging from $\beta = 1.5$ to $\beta = 3.0$. Throughout this paper, results are given in lattice units. Field configurations $U_\mu(x)$ distributed according to $\exp(-S[U])$ are generated and subsequently further decorrelated using the standard HMC algorithm. Details on the sampling procedure are described in \Cref{AppendixHybridMonteCarlo}. All expectation values given in this work are computed as averages of 100 samples for each $\beta$. No gauge-fixing is applied during sampling.

In order to understand the occurring structures in relation to ultraviolet fluctuations, we repeatedly compare with results from cooled samples. These are obtained by applying the standard Wilson flow~\cite{Luscher:2010iy}, which removes fluctuations by numerically solving a gradient flow equation that minimizes the Wilson action defined in \Cref{EqWilsonGaugeAction}, thereby smoothing the configurations. Further details of our cooling setup can be found in \Cref{AppendixWilsonFlow}.

\subsection{Confinement in Polyakov loops and their effective potential}\label{SecPolyakovIntro}

Winding around the periodic time direction, the Polyakov loop
\begin{equation}
    \Pp(\xx) = \mathscr{P} \prod_{\tau=1}^{N_\tau} U_4(\xx,\tau)
\,,
\end{equation}
where $\mathscr{P}$ indicates path ordering, provides a sensitive order parameter for confinement. Related to the free energy of static test quarks interacting via gluons, of particular interest is its trace~\cite{fukushima2017polyakov},
\begin{equation}
    P(\xx) = \frac{1}{2} \Tr \Pp(\xx)
\,,
\end{equation}
as well as the expectation value of its absolute volume average,
\begin{equation}
    L = \frac{1}{N_\sigma^3} \big\langle \big| \sum_{\mathbf{x} \in \Lambda_\sigma} P(\xx) \big|\big\rangle
\,.
\end{equation}
In order to account for the restoration of center symmetry for Polyakov loop observables in the statistical limit, for their evaluation we augment our Monte Carlo ensembles by adding duplicate samples with $P(\xx) \mapsto -P(\xx)$.

\begin{figure}
    \centering
	\includegraphics[scale = 0.69]{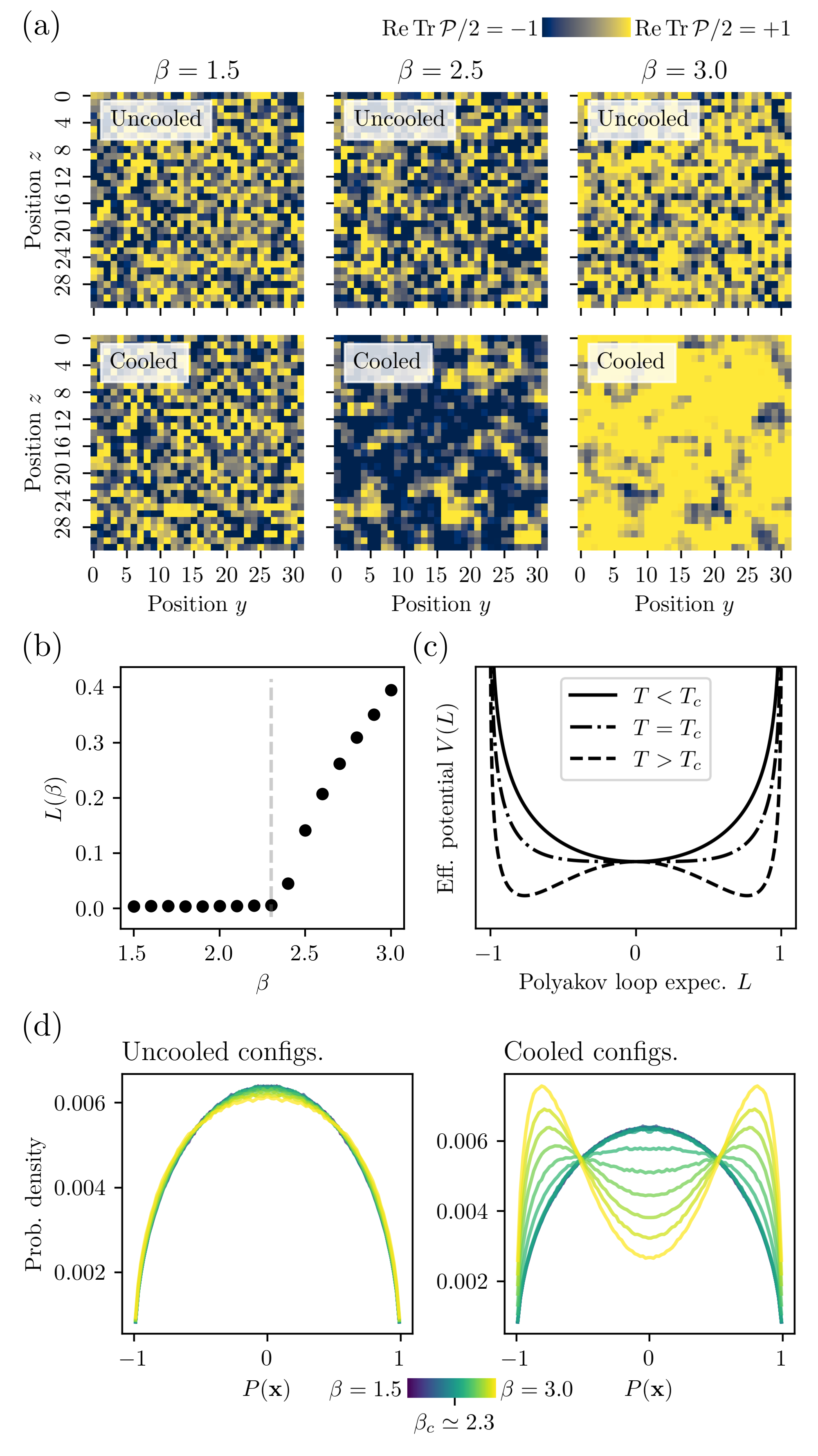}
	\caption{(a): Slices at constant $\xx^1$ of the Polyakov loop $P(\xx)$ without cooling applied (top row) and with cooling applied (bottom row), at different $\beta$. (b): Absolute value of the volume-averaged Polyakov loop expectation value versus $\beta$ with $\beta_c\simeq 2.3$ indicated (grey, dashed line). Errors are smaller than marker size. (c): Schematic, effective Polyakov loop potential in the confined phase (solid line), at the phase transition (dashed-dotted line) and in the deconfined phase (dashed line), derived from leading-order contributions of a strong coupling expansion and Haar measure (Vandermonde determinant) contributions \cite{fukushima2017polyakov}. (d): Probability densities of local Polyakov loop $P(\xx)$ values, for uncooled (left) and cooled configurations (right). Data is given in lattice units.}\label{FigPolyakovEffPotential}
\end{figure}

In \Cref{FigPolyakovEffPotential}(a) we display two-dimensional single-sample slices of $P(\xx)$ for three different values of $\beta$, of configurations with and without cooling applied. Throughout $\beta$-values, uncooled samples show many fluctuations on small length scales. For $\beta=2.5$ and $\beta=3.0$, a bias towards non-zero $P(\xx)$-values can be recognized. A comparison with cooled variants shows that for $\beta=1.5$, barely any structural changes can be observed. For $\beta=2.5$ and $\beta=3.0$, large domains of like-signed $P(\xx)$-values are visible after cooling.

Under a center transformation, $z\in Z(\mathrm{SU}(2)) = \{\pm 1\}$, the traced Polyakov loop transforms non-trivially as ${P(\xx)\mapsto z P(\xx)}$. Unbroken center symmetry requires $L=0$. If center symmetry is spontaneously broken, $L> 0$ is possible. This effect shows up in $L(\beta)$, see \Cref{FigPolyakovEffPotential}(b): $L(\beta) \approx 0$ below $\beta\simeq 2.3$ in our calculations, while for $\beta\gtrsim 2.3$ we find $L(\beta)>0$. This signals spontaneously broken center symmetry above $\beta \simeq 2.3$. We identify $\beta_c \simeq 2.3$ as the (approximate) critical inverse coupling squared\footnote{Note that the value of $\beta_c$ reported in the literature based on the computation of the Binder cumulant is slightly larger. A precise determination of $\beta_c$ is not the aim of this work, and we merely use the point where $L(\beta)$ becomes non-zero as a rough approximation.}, similar to previous works on critical properties of SU(2) lattice gauge theory on lattices of comparable size to ours \cite{engels1996critical, velytsky2008finite}. This explains the structures visible in \Cref{FigPolyakovEffPotential}(a): above $\beta_c$, spontaneous symmetry breaking is responsible for the formation of like-signed $P(\xx)$-domains, which in particular show up in cooled configurations.

All this can be attributed to a second-order phase transition manifesting in the effective Polyakov loop potential $V(L)$. We schematically display $V(L)$ in \Cref{FigPolyakovEffPotential}(c), where contributions from a leading-order large-coupling expansion and contributions from the SU(2) Haar measure (Vandermonde determinant) are taken into account \cite{fukushima2017polyakov}. Linearly mapping inverse couplings squared to temperatures, $V(L)$ has a global minimum at $L=0$ below $\beta_c$. $L=0$ corresponds to an infinite energy cost to excite a single static test quark. It thus becomes impossible to excite states which transform non-trivially under gauge transformations, indicating confinement. At $\beta_c$, the minimum becomes degenerate, leading for $\beta>\beta_c$ to two global minima at $L\neq 0$. Spatially, this is reflected by domain formation \cite{fortunato2000polyakov}. For uncooled and cooled configurations, the probability densities for local $P(\xx)$-values displayed in \Cref{FigPolyakovEffPotential}(d) agree with this phenomenology. Including quarks, similar probability densities have been computed in \cite{langfeld2013two}; the phase transition smears out to a crossover at finite baryon densities.

Correlations of multiple traced Polyakov loops are related to the free energy of static multi-quark configurations and similarly make up interpretable confinement order parameters \cite{fukushima2017polyakov}. In \Cref{AppendixPolyakovCorrels}, we discuss two-point function results for Polyakov loop correlations computed from the lattice data.

Non-trivial topology is related to non-trivial holonomies in the Polyakov loop. For $|\xx| \to \infty$, the Polyakov loop is diagonalizable with
\begin{equation}
\lim_{|\xx|\to\infty} \Pp(\xx) = \left(\begin{matrix}
\exp(2\pi i \mu_1) & 0\\
0 & \exp(2\pi i \mu_2)
\end{matrix}\right)\,,
\end{equation}
with $\mu_1\leq \mu_2$ and $\mu_1+\mu_2=0$. The holonomies $\mu_1,\mu_2$ are related to the masses $8\pi^2 (\mu_2-\mu_1)$ and $8\pi^2 (1+\mu_1 - \mu_2)$ of SU(2) instanton-dyons, two of which form a KvBLL caloron \cite{kraan1998periodic, kraan1998monopole, lee19982}. Instanton-dyons as well as calorons are (anti-)self-dual solutions to the classical Yang-Mills equations: $F_{\mu\nu}(x) = \pm\tilde{F}_{\mu\nu}(x)$, i.e., $\EE(x) = \pm \BB(x)$. Holonomy potentials computed from instanton-dyon ensembles can approximately account for the Polyakov loop effective potential $V(L)$ \cite{diakonov2009topology, lopez2018confinement}.

Without cooling, calorons and instanton-dyons are specific field configurations difficult to detect in lattice data. However, the cooling process also unavoidably removes relevant physical information. This prompts the question whether persistent homology is able to detect meaningful signatures of the underlying topology in the raw data.

\section{Geometric structures in Polyakov loops from persistent homology}\label{SecPolyakPersHom}
Persistent homology provides a means to algorithmically deduce topological structures from potentially noisy numerical data, including with \emph{persistence} a measure of their dominance. In this section, we utilize persistent homology of different types of filtrations constructed from Polyakov loops and related algebra fields in order to unravel relevant (de-)confinement features. The latter include the formation of spatial lumps of topological density instead of extended, string-like configurations, accompanied by probabilistic evidence for the occurance of calorons and instanton-dyons. We stress that all of the involved constructions are gauge-invariant without a priori assumptions on the type of excitations under study.

We begin in \Cref{SecIntroPersHom} with an intuitive introduction to cubical complexes and persistent homology. In \Cref{SecPolyakTopDens} we discuss persistent homology results of the sublevel set filtration of topological densities computed from Polyakov loops, first introducing the latter. \Cref{SecAngleDiffFiltration} is devoted to signals of the confinement phase transition in the so-called angle difference filtration applied to Polyakov loop algebra element norms.

\subsection{A primer on persistent homology via sublevel sets of the traced Polyakov loop}\label{SecIntroPersHom}

We introduce concepts of persistent homology with the example of the traced Polyakov loop field ${P(\xx)=(1/2) \Tr\Pp(\xx)}$ and its sublevel set filtration of cubical complexes\footnote{The values of $P(\xx)$ scattering evenly around zero, we expect similar results for superlevel instead of sublevel sets.}. Technical details of mathematical constructions are given in \Cref{SecAppendixMathPersDiags}. For more elaborate introductions to persistent homology and computational topology, we refer to the literature \cite{carlsson2009topology, otter2017roadmap, edelsbrunner2022computational}, similarly for general introductions to algebraic topology \cite{may1999concise, munkres2018elements}.

The \emph{sublevel sets} $M_P(\nu)$ of $P(\xx)$ displayed in \Cref{FigPolyakovEffPotential}(a) are subsets of the spatial lattice,
\begin{equation}
M_{P}(\nu) := P^{-1}(-\infty,\nu] = \{\xx\,|\, P(\xx)\leq \nu\}\,.
\end{equation}
Superlevel sets are defined as 
\begin{equation}
N_P(\nu) := P^{-1}[\nu,\infty) = \{\xx\,|\,P(\xx)\geq \nu\}\,.
\end{equation}
The lattice Polyakov loop $P(\xx)$ approximates the continuum Polyakov loop in a cube\footnote{To clarify notations:\\${\xx+[-1/2,1/2]^3 = \{\yy\in \rr^3\,|\, \yy-\xx\in [-1/2,1/2]^3\}}$.} $\xx + [-1/2,1/2]^3$. The (point set) topology of $M_P(\nu)$ and $N_P(\nu)$ merely amounting to point counting, we seek to construct spaces reflecting the cubical structures while still corresponding to $M_P(\nu)$, $N_P(\nu)$.

\subsubsection{Filtered cubical complexes}\label{SecFilteredCubicalComplexes}
Particularly suitable for the algorithmic computation of topological descriptors of pixelized data are \emph{cubical complexes}. Let $\Cc$ be the full cubical complex, consisting of one cube $\xx+[-1/2,1/2]^3$ of top dimension (here, three) for each spatial lattice point $\xx$. Every 3-cube comes with all its faces, edges and vertices. In fact, it is a defining property of cubical complexes to be closed under taking boundaries of any of its top- and lower-dimensional cubes. The boundary of a 3-cube is the union of all its six faces, the boundary of a 2-cube (face) is the union of its four boundary edges, the boundary of a 1-cube (edge) consists of its two endpoints, and the boundary of a 0-cube (point) is empty. The boundary operator $\partial$ provides the map from a cube to its boundary.

How can we equip the full cubical complex $\Cc$ with the information contained in $P$? We inductively construct a suitable map ${\tilde{P}:\Cc\to \rr}$. Any 3-cube $C\in \Cc$ has a unique lattice point $\xx$ at its center. We set $\tilde{P}(C):=P(\xx)$. Any 2-cube $D\in\Cc$ is contained in the boundary of two 3-cubes. For all 2-cubes $D\in \Cc$ we set
\begin{equation}\label{EqLowerStarFiltrationConstruction}
\tilde{P}(D):=\min \{\tilde{P}(C)\,|\, D\in \partial C, C\in\Cc\,\text{3-cube}\}\,.
\end{equation}
Similarly, any 1-cube is contained in the boundaries of four 2-cubes, any 0-cube is contained in the boundaries of six 1-cubes. \Cref{EqLowerStarFiltrationConstruction} is applied to all 1-cubes instead of 2-cubes with values induced from 2-cubes, and finally to all 0-cubes with values induced from 1-cubes, until $\tilde{P}$ is defined on all cubes of all dimensions in $\Cc$. Sublevel sets of $\tilde{P}$,
\begin{equation}
\Cc_P(\nu):=\tilde{P}^{-1}(-\infty,\nu] = \{C\in \Cc\,|\, \tilde{P}(C)\leq \nu\}\,,
\end{equation}
are closed under taking boundaries, constituting cubical complexes which correspond to pixelizations of the lattice sublevel sets $M_P(\nu)$. We define superlevel set cubical complexes as
\begin{equation}
\Dd_P(\nu):= \Cc_{-P}(-\nu)\,,
\end{equation}
corresponding to the lattice superlevel sets $N_P(\nu)$. All these constructions work analogously in higher dimensions.

The sublevel set cubical complexes $\Cc_P(\nu)$ form a \emph{filtration} of cubical complexes,
\begin{equation}
\Cc_P(\nu)\subseteq \Cc_P(\mu),\;\text{whenever}\; \nu\leq \mu\,.
\end{equation}
With its construction, this filtration is called the \emph{lower-star filtration}.
Analogously, a filtration of superlevel set cubical complexes occurs,
\begin{equation}
\Dd_P(\nu) \supseteq \Dd_P(\mu),\;\text{whenever}\;\nu \leq \mu\,.
\end{equation}
The lattice $\Lambda$ consisting of finitely many points, these filtrations consist of only finitely many distinct complexes.

\subsubsection{Persistent homology: holes in complexes}

\begin{figure}
    \centering
	\includegraphics[scale = 0.69]{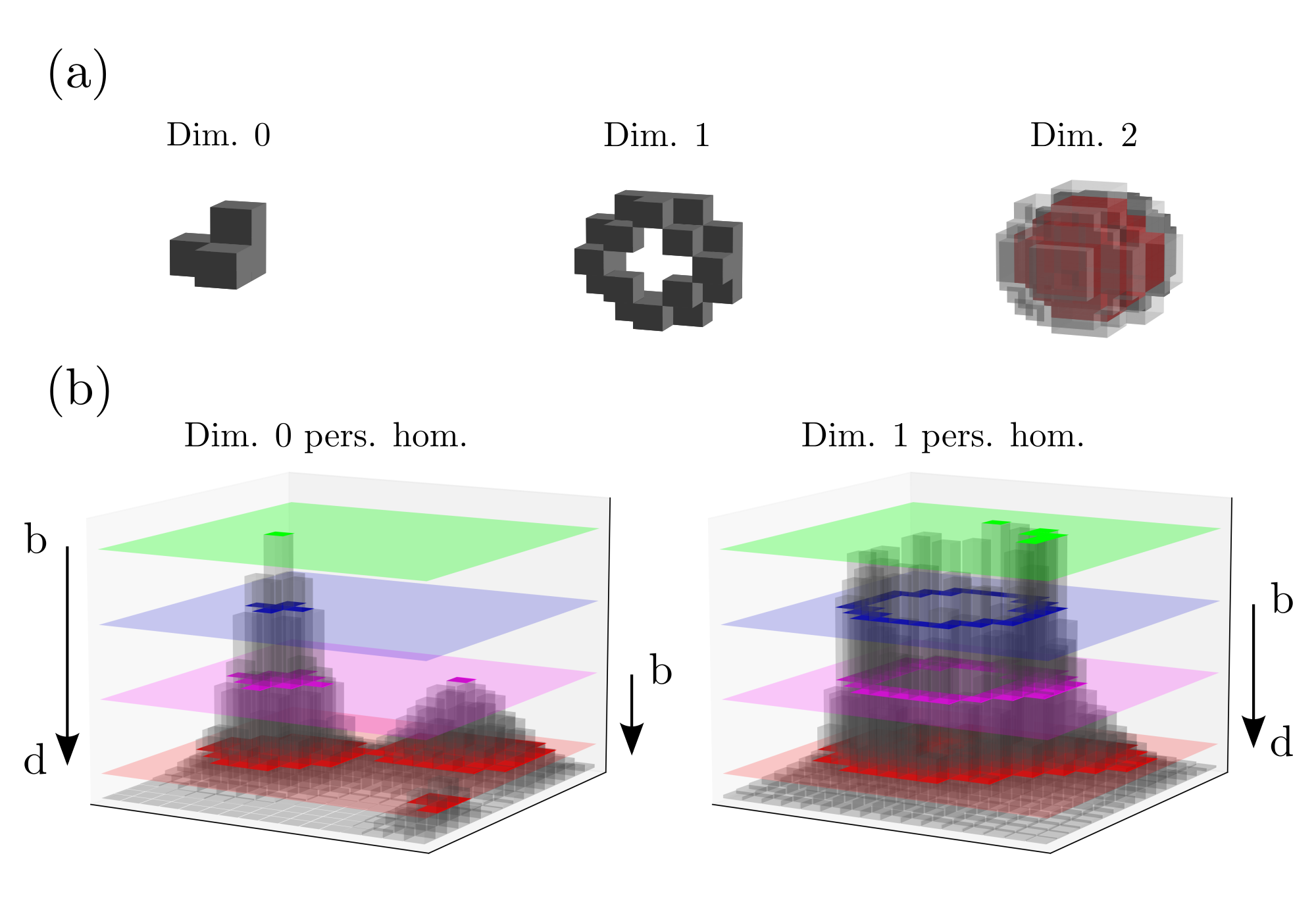}
	\caption{(a): Cubical complexes giving rise to homology classes of dimensions zero to two from left to right. The enclosed volume in dimension two is indicated in red. (b): Schematically, dimension zero and dimension one persistent homology classes of superlevel sets of a function with two-dimensional domain. Exemplary superlevel set cubical complexes are indicated by different colors; birth and death of corresponding persistent homology classes are indicated by $b$ and $d$, respectively.}\label{FigPersHomSchematically}
\end{figure}

Generically, the cubical complexes $\Cc_P(\nu)$ do not contain a cube for every spatial lattice point. Holes appear, described by homology groups. Their elements, homology classes, are constructed comparably to homotopy classes and capture similar topological information \cite{may1999concise}, but are algebraically much better accessible. In particular, homology classes are also homotopy-invariant, i.e., continuous deformations of holes leave homology classes invariant. For an impression of low-dimensional homology classes, we refer to \Cref{FigPersHomSchematically}(a). Connected components are described by dimension zero homology classes. Planar-like holes, circumscribed by a circle which cannot be continuously deformed into a point within the cubical complex, belong to the dimension one homology group. The dimension two homology group captures  enclosed volumes, described by a 2-sphere. Higher-dimensional homology classes correspond to analogous higher-dimensional holes.

As in cubical complex filtrations the filtration parameter $\nu$ is swept through, homology classes may be born and die again. \emph{Persistent homology} captures this. In \Cref{FigPersHomSchematically}(b) we indicate two scenarios for superlevel sets of exemplary functions on a surface. The left function shows three distinct peaks. For $\nu$ larger than the maximum value of the highest peak the cubical complex is empty with trivial homology. As $\nu$ is lowered to exactly the latter value (indicated in \Cref{FigPersHomSchematically}(b), left panel, by the green plane), a zero-dimensional homology class is born, with \emph{birth} $b=\nu$. Lowering $\nu$ further (blue plane), the single 2-cube turns into an accumulation of 2-cubes, nothing changing in homology. At the $\nu$-value indicated by the pink plane, a second zero-dimensional homology class is born with the second peak showing up in the complex. Again, for the first homology class nothing changes. At the $\nu$-value indicated by the red plane, the two accumulations of 2-cubes merge into one. The first homology class dies with $d=\nu$ its \emph{death}. We call $p=d-b$ its \emph{persistence}. The second homology class dies later upon merging with the third peak (not indicated). Turning to the right function in \Cref{FigPersHomSchematically}(b), for $\nu$-values larger than the one indicated by the blue plane, different dimension-zero homology classes are born and die upon merging with each other. For $\nu$ corresponding to the blue plane, a circular structure surrounding a hole appears in the corresponding complex: a one-dimensional homology class is born. For $\nu$ between the pink and red planes, the hole disappears, getting fully filled by 2-cubes. The homology class dies.

In higher dimensions, dimension zero homology classes can still be imagined as independent connected components. Dimension one homology classes correspond to structures such as approximate circles or empty tori surrounded by cubes. Dimension two homology classes are empty 3-volumes.

The \emph{$\ell$-th persistence diagram} $\Dgm_\ell(\Cc_P)$ consists of birth-death pairs $(b,d)$, one for each independent homology class of dimension $\ell$ in a given filtration such as $\Cc_P$. \emph{Betti numbers} $\beta_\ell(\nu)$ specify the number of independent $\ell$-dimensional homology classes in $\Cc_P(\nu)$. They can be obtained from $\Dgm_\ell(\Cc_P)$ as
\begin{equation}
\beta_\ell(\nu) = \#\{(b,d)\in\Dgm_\ell(\Cc_P)\,|\, b\leq \nu < d\}\,.
\end{equation}
From $\Dgm_\ell(\Cc_P)$ we may also obtain statistics such as the number of independent homology classes with a given birth $b$, $\Bb_\ell(b)$, or the number of independent homology classes with a given persistence $p$, $\Pp_\ell(p)$. Note that homology classes can have infinite persistence. For instance, the full cubical complex $\Cc$ is one connected component. Also, periodic boundaries turn the full complex $\Cc$ into a 3-torus which has three independent dimension one and three independent dimension two homology classes, and even a one-dimensional homology group in dimension three.

Statistically evaluating expectation values, throughout this work persistent homology quantifiers are computed from individual samples and subsequently averaged. The investigated persistent homology descriptors can be consistently defined in a statistical setting with limit theorems for large-volume asymptotics existing \cite{hiraoka2018limit, spitz2020self}. Well-defined thermodynamic limits actually require the latter.

Importantly, persistence diagrams are stable against small perturbations of the input function $P$ \cite{cohen2007stability, cohen2010lipschitz, bauer2013induced}, which facilitates the numerics. We compute persistent homology of cubical complexes with periodic boundary conditions using the Python and C++ interfaces of the versatile computational topology library GUDHI \cite{maria2014gudhi}.

\begin{figure*}[t]
    \centering
	\includegraphics[scale = 0.69]{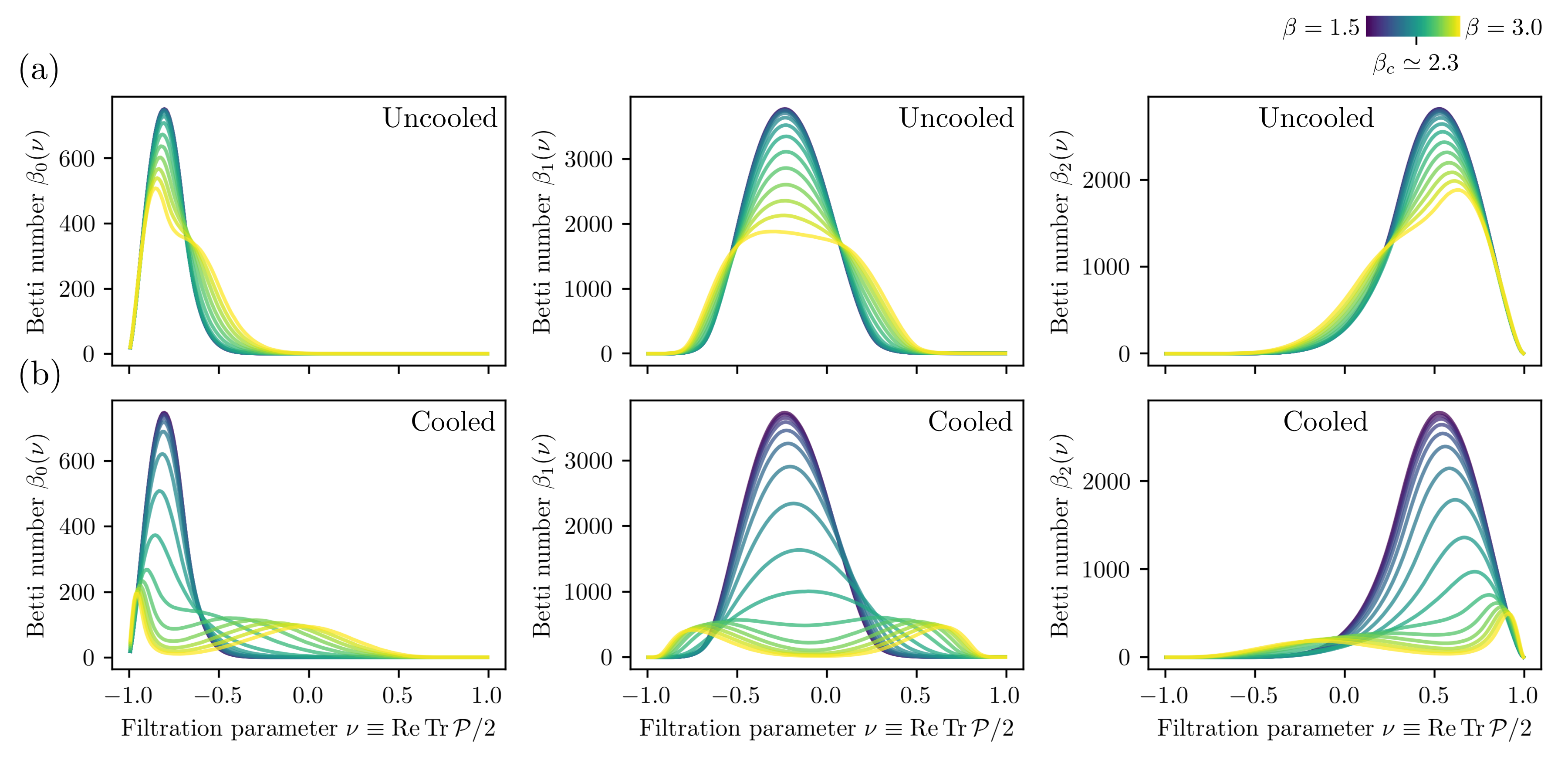}
	\caption{Betti number distributions of dimensions zero to two for the sublevel set filtration of $P(\xx)=\Tr\Pp(\xx)/2$ for (a) uncooled and (b) cooled configurations.}\label{FigReTrPBetti}
\end{figure*}

\subsubsection{Traced Polyakov loop results}\label{SecReTrPBetti}
Numerical Betti number distributions of different dimensions of the sublevel set filtration $\Cc_P(\nu)$ are displayed in \Cref{FigReTrPBetti} for configurations without and with cooling applied, for all inverse couplings squared from $\beta=1.5$ to $\beta=3.0$. All Betti number distributions show a distinct peak, whose position consistently shifts to larger filtration parameters $\nu$ with increasing homology class dimensions. This is characteristic to sublevel set filtrations: lowest in the filtration, connected components form in order to merge into extended circular structures at larger filtration parameters, similar to \Cref{FigPersHomSchematically}(b). Typically, enclosed voids form out of pitted, hollow networks of cubes with many dimension one homology classes dying in order to form a dimension two homology class. This behavior usually continues to higher dimensions.

Without cooling, we observe from \Cref{FigReTrPBetti}(a) that Betti number distributions approximately lay on top of each other for $\beta\lesssim \beta_c \simeq 2.3$. For $\beta\gtrsim\beta_c$ and throughout dimensions, peaks diminish in height with $\beta$ and broaden. In particular, in dimensions zero and two, an additional bump occurs on the peak side pointing towards filtration parameter zero. As seen in \Cref{FigReTrPBetti}(b), cooling enhances these effects. The peaks already decrease in height for $\beta > 1.8$, with moderately constant peak positions up to $\beta \simeq 2.2$. For $\beta \gtrsim 2.3$, the peak splits up in two. Qualitatively, dimension zero and two Betti number distributions seem to be mirrored at $\nu=0$.

These observations can be understood from the effective Polyakov loop potential $V(L)$, see \Cref{FigPolyakovEffPotential}(c). Below $\beta_c\simeq 2.3$, the $\zz_2$ center symmetry of $L$ is unbroken: the distribution of local $P(\xx)$-values is symmetric around zero and approximately constant, see \Cref{FigPolyakovEffPotential}(d). The constancy of uncooled $\beta_\ell(\nu)$ for $\beta\lesssim \beta_c$ is a manifestation of this. Also, the qualitative mirroring between dimensions zero and two reflects this: in sublevel set filtrations, a minimum shows up as a dimension zero homology class. A maximum shows up as a dimension two homology class. Both are expected to occur comparably likely if the $\zz_2$ symmetry is unbroken.

Above $\beta_c$, the center symmetry is spontaneously broken, see \Cref{FigPolyakovEffPotential}(a): individual samples acquire non-zero volume averages $L$. Center symmetry is restored only in the statistical limit. In Betti number distributions, the appearance of additional bumps without cooling or two peaks with cooling resembles the spontaneous symmetry breaking behavior and statistical restoration of center symmetry. The decrease in the peak's heights can be understood from a homogenization of $P(\xx)$ above $\beta_c$: structures become fewer. For cooled configurations, effects are enhanced due to less ultraviolet fluctuations.

Domains of like-signed Polyakov loops forming, the behavior is consistent with a percolation interpretation of deconfinement \cite{fortunato2000polyakov}. Beyond that, can we identify topological excitations?

\subsection{Sublevel sets of topological densities from Polyakov loops}\label{SecPolyakTopDens}
Topological excitations such as calorons are characterized by non-trivial topological densities. However, in typical Monte Carlo samples, topological densities contain strong signatures of ultraviolet fluctuations and of the lattice discretization. Effectively averaging temporal fluctuations, Polyakov loops reveal less such fluctuations. We consider in this section the sublevel set filtration of the Polyakov loop topological density on the lattice,
\begin{align}
q_\Pp(\xx):=&\;\frac{1}{24\pi^2}\varepsilon_{ijk}\Tr[(\Pp^{-1}(\xx)\partial_i \Pp(\xx))\nonumber\\[1ex]
&\quad \times (\Pp^{-1}(\xx)\partial_j \Pp(\xx))(\Pp^{-1}(\xx)\partial_k \Pp(\xx))]\,.
\end{align}
For comparison, we discuss the superlevel set filtration of the usual topological density $q\sim \Tr\EE\cdot\BB$ in \Cref{SecEBTopDens}.

Indeed, the nomenclature is justified. In a theory with continuous space-time and periodic boundary conditions, i.e., with space-time the 4-torus $T^4$, the Polyakov loop $\Pp$ is a map from the 3-torus $T^3$ to ${\mathrm{SU}(2)\cong S^3}$. Under fairly general assumptions, its winding number may be computed as
\begin{equation}
Q_{\mathrm{top}} = \frac{1}{32\pi^2}\int_{T^4}\varepsilon_{\alpha\beta\mu\nu}\Tr F_{\alpha\beta} F_{\mu\nu}=\int_{B_4} q_{\Pp}\,,
\end{equation}
where $\varepsilon_{\alpha\beta\mu\nu}$ is the Levi-Civita symbol in four dimensions and ${B_4=\{(\xx,\tau)\in T^4\,|\, \tau=0\}}$ \cite{ford1998monopoles}. For details on this rewriting, we refer to \Cref{AppendixTopDensFromPolyak}.

\begin{figure*}
    \centering
	\includegraphics[scale = 0.69]{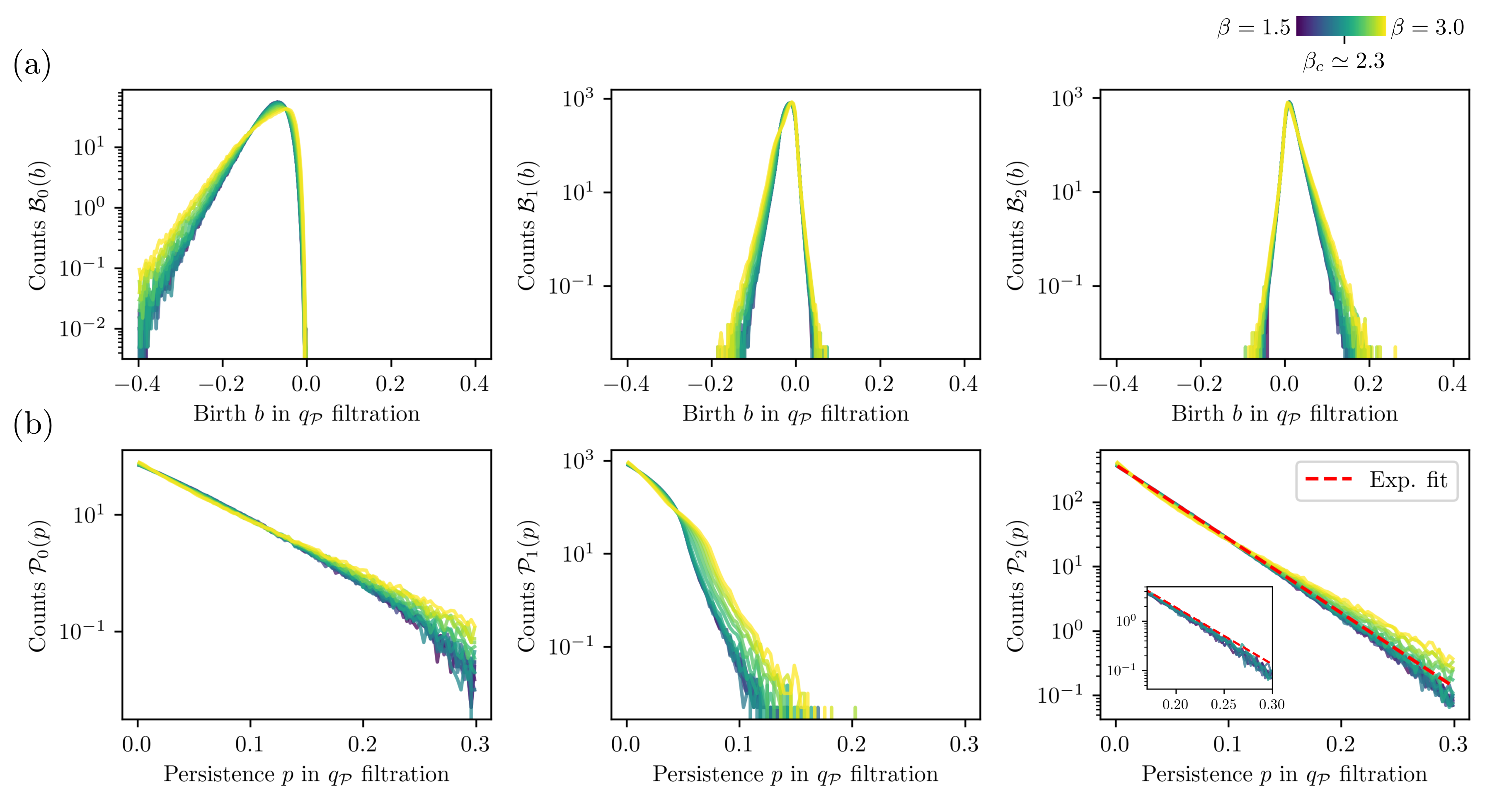}
	\caption{Homological quantifiers of the sublevel set filtration of the Polyakov loop topological density $q_{\mathcal{P}}(\xx)$. (a): Birth distributions for dimensions zero to two. (b): Persistence distributions for dimensions zero to two. The exponential fit of the $\beta=1.5$ persistence distribution reveals $\Pp_2(p)\sim \exp(-sp)$ with $s=26.51\pm 0.04$. Inset shows large-persistence, dimension two persistence distributions for inverse couplings up to $\beta_c$. No cooling has been applied in all data shown in this Figure, though cooling leaves the data shown approximately invariant for $\beta \lesssim \beta_c\simeq 2.3$. Results are given in lattice units.}\label{FigPolyakovTopDens}
\end{figure*}

In \Cref{FigPolyakovTopDens}, we display birth and persistence distributions of the sublevel set filtration of $q_\Pp$ for uncooled configurations. 
Cooling barely has any effect on the shown homological descriptors below $\beta_c$ and enhances the occurring trends above $\beta_c$, see \Cref{AppendixTopDensFromPolyak}. We deduce that topologically non-trivial excitations are mostly due to (near-)classical configurations. Similar to the traced Polyakov loop, the birth distributions in \Cref{FigPolyakovTopDens} reveal a single distinct peak for each $\beta$. Birth distributions are constant for $\beta\lesssim \beta_c$. Above $\beta_c$, distributions broaden for increasing $\beta$. Dimension zero birth distributions have support below birth $b=0$, dimension one around birth $b\approx 0$, and dimension two with a bias towards birth $b>0$. 

The dimension zero and dimension two persistence distributions shown in \Cref{FigPolyakovTopDens}(b) are almost identical in shape. For most $\beta$-values, these distributions follow nearly exponential behavior, approximately constant for $\beta\lesssim \beta_c$. An exponential fit of the $\beta=1.5$ dimension two persistence distribution reveals $\Pp_2(p)\sim \exp(-sp)$ with $s=26.51\pm 0.04$. The dimension one persistence distribution does not show comparable behavior for persistences $p\gtrsim 0.02$ and quickly declines. Up to $p\approx 0.02$ it looks similar to dimensions zero and two.

We may interpret the similarity of persistence distributions in dimensions zero and two analogously to the qualitative similarity of Betti numbers of traced Polyakov loops in \Cref{FigReTrPBetti}(b): $q_\Pp$ being statistically symmetric around zero, minima and maxima occur comparably likely. While the former show up as dimension zero homology classes in the $q_\Pp$-sublevel set filtration, the latter give rise to dimension two homology classes. $q_\Pp(\xx)$ is governed by local accumulations of non-zero topological density. Dimension one homology classes originate primarily from (thermal) noise, explaining the small-persistence support of $\Pp_1(p)$.

For well-separated peaks, the persistence $p$ of dimension zero and dimension two homology classes provides a quantifier of their dominance.
In \cite{larsen2016classical} it has been argued that for an instanton dyon-antidyon pair [of type $M\bar{M}$ or $L\bar{L}$ in SU(2)] at large separation $r$, the due to time-independence effectively three-dimensional action $S_3$,
\begin{equation}
S = \frac{1}{g^2}\int_0^{1/T} d\tau S_3\,,
\end{equation}
behaves as
\begin{equation}\label{EqS3}
S_3 = 8\pi v + (m_1m_2-e_1e_2)\frac{4\pi}{r}\,,
\end{equation}
with dyon magnetic charges $m_1$, $m_2$ and electric charges $e_1$, $e_2$. The holonomy parameter $v$ originates from ${A_4^a(x\to\infty)\to v \hat{r}_a}$ for a given $\mathfrak{su}(2)$ direction vector $\hat{r}_a$ in the instanton-dyon parametrization used \cite{larsen2016classical}. If we take $T =1/g^2$, then for time-independent configurations $S_3 = S$. For self-dual excitations, $S$ further equates to $Q_{\mathrm{top}}=\int_{\xx} q_\Pp(\xx)$. One is tempted to observe that, at leading order in $1/r$, ${S_3/v \simeq 8\pi \simeq 25.1}$ is close to the fitted value of $s\simeq 26.5$ for the dimension two persistence distribution in the confining phase. Effects of the lattice discretization, thermal noise, contributions from configurations with more than two dyons\footnote{For in total $n$ instanton-(anti-)dyons it is expected that the $8\pi v$ contribution is to be replaced by $4\pi n v$ \cite{diakonov2009topology}.}, and in general $g^2T\neq 1$ are expected. Nonetheless, we find that in the confining phase, persistence distributions of $q_\Pp$ resemble the exponential behavior of the (semi-)classical instanton-dyon occurrence probability $\sim \exp(-S)$. The behavior of $\Pp_\ell(p)$ above $\beta_c$ shows growing deviations from this. Other topological excitations such as domain walls can show up in $q_\Pp$ configurations above $\beta_c$ due to spontaneous center symmetry breaking. It is suggestive that growing persistences above $\beta_c$ are at least partially triggered by these topological structures.

\begin{figure*}[t]
    \centering
	\includegraphics[scale = 0.69]{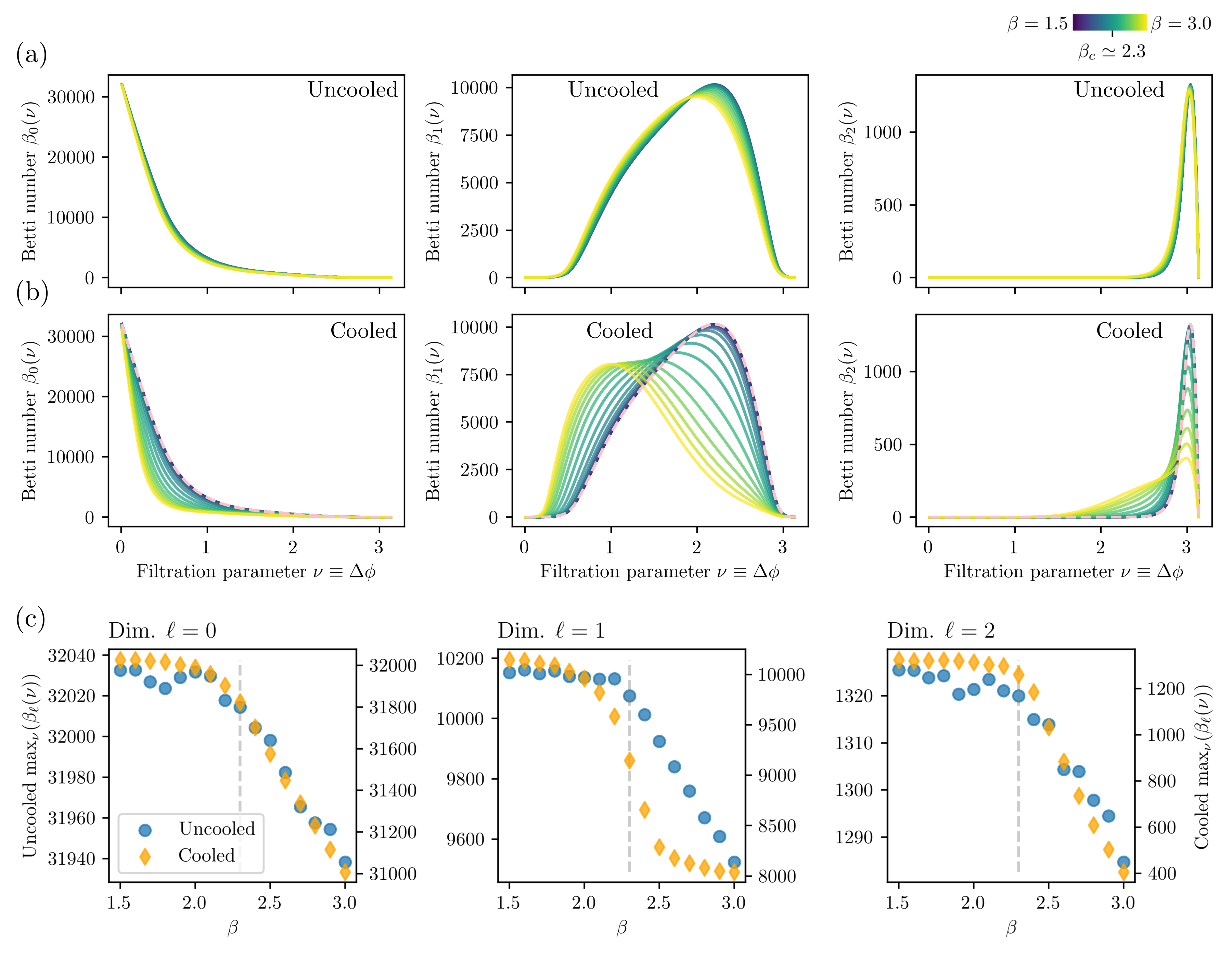}
	\caption{Homological quantifiers of the angle difference filtration constructed from $\phi(\xx) = \arg (\Tr\Pp(\xx)/2)$. (a): Betti number distributions of uncooled configurations for dimensions zero to two. (b): Betti number distributions of cooled configurations for dimensions zero to two, the pink dashed line indicating the uncooled Betti number distribution for $\beta=1.5$. (c) Maxima of Betti number distributions of dimensions zero to two versus $\beta$ for uncooled (left axes) and cooled configurations (right axes).}\label{FigAngleDiff}
\end{figure*}

\begin{figure}[t]
    \centering
	\includegraphics[scale = 0.69]{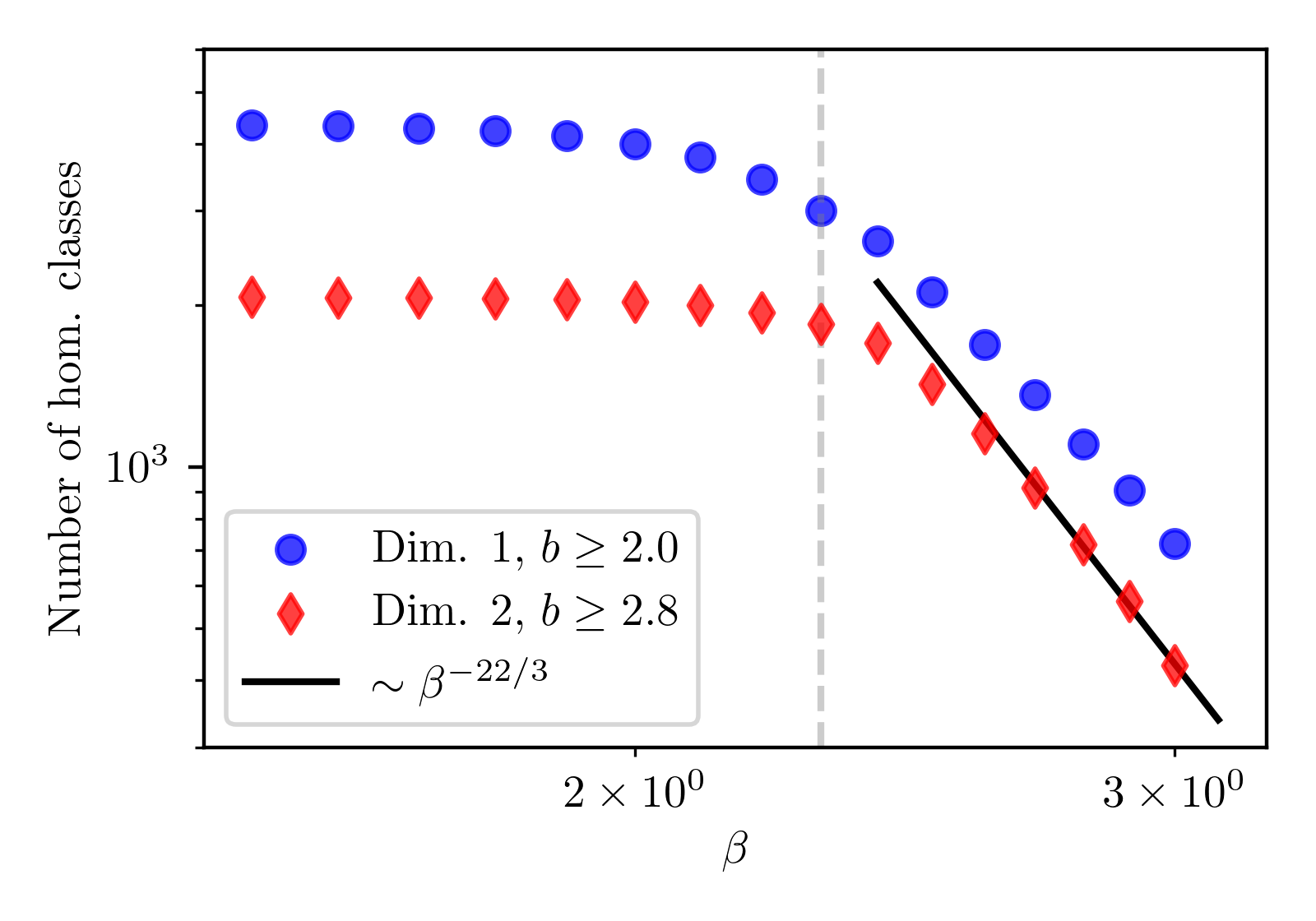}
	\caption{Number of homology classes in the angle difference filtration constructed from $\phi(\xx) = \arg (\Tr\Pp(\xx)/2)$ with birth $b\geq 2.0$ for dimension one and $b\geq 2.8$ for dimension two, compared to the semi-classical one-loop instanton appearance probability scaling behavior $\sim \beta^{-22/3}$.}\label{FigAngleDiffNoHomClasses}
\end{figure}

\subsection{Angle-difference filtration of local Polyakov holonomies}\label{SecAngleDiffFiltration}

The SU(2)-valued Polyakov loop can be written in terms of an algebra field as 
\begin{align}\label{eq:AlgPol}
{\Pp(\xx) = \exp(i\phi^a(\xx)T^a})\,,
\end{align}
with $T^a = \sigma^a/2$ the Hermitian $\mathfrak{su}(2)$ generators given in terms of the Pauli matrices $\sigma^a$. A $2\pi$-periodic scalar field can be defined as half the norm of the Polyakov loop Lie algebra components,
\begin{equation}
\phi(\xx)= \frac{1}{2}\sqrt{(\phi^1(\xx))^2 + (\phi^2(\xx))^2 + (\phi^3(\xx))^2}\,.
\end{equation}
Then, $P(\xx)=\Tr \Pp(\xx)/2 = \cos \phi(\xx)$. If the traced Polyakov loop $P(\xx)$ locally changes, this can give rise to non-trivial spatial structures occurring in $\phi(\xx)$.

In \cite{sale2022quantitative}, a filtration has been constructed, that is only sensitive to local differences of periodic fields. The filtration differs from the lower-star filtration discussed so far in that function values are assigned to edges instead of top-dimensional cubes. Specifically, with $\Cc$ again the full 3-dimensional cubical complex as in \Cref{SecFilteredCubicalComplexes}, we construct from $\phi(\xx)$ a map $\tilde{\phi}:\Cc\to [0,\pi]$, whose sublevel sets then form cubical subcomplexes of $\Cc$. Lattice points $\xx\in \Lambda_\sigma$ are mapped to vertices of the full complex $\Cc$, defining\footnote{Strictly speaking, we here choose the 3-cubes of $\Cc$ to be $\xx+[0,1]^3$ for lattice points $\xx\in\Lambda_\sigma$.} $\tilde{\phi}(\{\xx\}):=0$. With $\xx$ and $\yy$ nearest neighbors in $\Lambda_\sigma$, we set
\begin{equation}
\Delta\phi(\xx,\yy):= \min \{|\phi(\xx)-\phi(\yy)|,2\pi - |\phi(\xx)-\phi(\yy)|\}\,,
\end{equation}
and define $\tilde{\phi}(\{\xx,\yy\}):=\Delta\phi(\xx,\yy)$ with $\{\xx,\yy\}$ the edge connecting $\xx$ with $\yy$. We extend to higher-dimensional cubes via the \emph{upper-star filtration}, i.e., we induce values from lower-dimensional cubes,
\begin{equation}
\tilde{\phi}(C):=\max\{\tilde{\phi}(D)\,|\,D\in \partial C\}\,,
\end{equation}
and apply this construction inductively until $\tilde{\phi}$ is defined on all $\Cc$. Sublevel sets of $\tilde{\phi}$ yield the \emph{angle-difference filtration},
\begin{equation}
\tilde{\phi}^{-1}[0,\nu] \subseteq \tilde{\phi}^{-1}[0,\mu],\;\mathrm{whenever}\; \nu\leq \mu\,,
\end{equation}
whose persistent homology we shall investigate.

By construction, the angle-difference filtration of $\phi(\xx)$ does not contain information on the volume-averaged traced Polyakov loop expectation value $L$. Indeed, we expand $L$ in terms of $\phi(\xx)$:
\begin{align}\nonumber 
    L =&\; \frac{1}{N_\sigma^3} \big\langle \big|\sum_{\xx\in\Lambda_\sigma} \cos (\phi(\xx))\big|\big\rangle\\[1ex]
    =&\;  \frac{1}{N_\sigma^3} \big\langle \big| \sum_{n=0}^\infty \frac{(-1)^n}{(2n)!} \sum_{\xx\in\Lambda_\sigma} \phi^{2n}(\xx)\big|\big\rangle\,,
\end{align}
i.e., only volume-averages of even powers of $\phi$  enter $L$. The spatial average $\sum_{\xx\in \Lambda_\sigma}\phi(\xx)/N_\sigma^3$ does not enter the angle-difference filtration of the holonomy Lie algebra field $\phi$ directly. In addition, the angle-difference filtration of $\phi$ is by construction center symmetric ($\phi\mapsto \phi+\pi$).

Betti number distributions of the angle-difference filtration for uncooled and cooled configurations are displayed in \Cref{FigAngleDiff}(a) and (b), respectively. Points entering as independent connected components at filtration parameter $\nu=0$ and merging at $\nu = \Delta\phi(\xx,\yy)$ into edges, by construction dimension zero Betti numbers monotonously decrease with growing filtration parameters and contain information on the statistics of nearest-neighbor $\Delta\phi(\xx,\yy)$-values.
Throughout dimensions, Betti number distributions of uncooled configurations are constant for $\beta\lesssim \beta_c$. For $\beta$ near 1.5, cooling leaves the Betti number distributions invariant (see the pink, dashed line in \Cref{FigAngleDiff}(b)). Dimension one Betti numbers show a distinct peak around $\nu\equiv\Delta\phi \simeq 2.2$ for $\beta\lesssim \beta_c$, slightly wandering towards smaller filtration parameters with increasing $\beta$. This behavior is pronounced for cooled configurations and starts off at smaller $\beta$ already. Dimension two Betti numbers of uncooled configurations show a prominent peak around $\nu\simeq 3.0$. After cooling, it decreases vastly in height for increasing $\beta$. A population of dimension two homology classes emerges at smaller $\nu$ for larger $\beta$.

The above observations can be understood as follows: Below $\beta_c$, Polyakov loop samples are dominated by vast fluctuations between $\approx -1$ and $\approx +1$ on tiny length scales, see \Cref{FigPolyakovEffPotential}(a). The formation of dimension two persistent homology classes in the angle-difference filtration thus requires large phase jumps of order $\pi$ to occur everywhere around local minima or maxima, thus occurring in the angle-difference filtration below $\beta_c$ primarily for $\nu\approx \pi$. As discussed in \Cref{SecReTrPBetti}, above $\beta_c$, extended domains of like-signed $P(\xx)$-values form, however without cooling strongly overlaid by thermal fluctuations. Fewer dimension two homology classes corresponding to extended domains fit into the given lattice volume compared to those originating from single-pixel fluctuations, explaining the peak decline in $\beta_2(\nu)$ in particular for cooled configurations. On top of the like-signed domains forming above $\beta_c$, fluctuations occur. The population of dimension two homology classes emerging for $\beta \gtrsim \beta_c$ below the $\nu\simeq 3.0$-peak may be understood as a signal of these. 

Dimension one Betti numbers show the strongest $\beta$-dependence. In \cite{sale2022quantitative}, the angle-difference filtration has been employed to uncover the behavior of vortices in two spatial dimensions. In three dimensions, these would show up as closed vortex lines, which would manifest as signals in dimension one Polyakov topological densities. It has been a key finding of \Cref{SecPolyakTopDens} that $q_\Pp$-sublevel sets barely show such signatures, but instead predominantly give rise to local lumps of topological density. We thus expect that the strong $\beta$-dependence of dimension one Betti numbers originates from locally large phase gradients spreading through space around like-signed Polyakov loop regions with fluctuations on top. Then, the angle-difference filtration yields at intermediate filtration parameters randomized scaffold-like cubical complexes with a variety of one-dimensional homology classes occurring. This is qualitatively in accordance with space-filling instanton-dyon positions in models of instanton-dyon ensembles \cite{diakonov2009topology, larsen2015interacting}.

Displayed in \Cref{FigAngleDiff}(c), maxima of Betti numbers of uncooled configurations show a clear kink around $\beta_c$, thus providing an order parameter for the confinement-deconfinement phase transition. 

In \Cref{FigAngleDiffNoHomClasses}, we display the number of homology classes with high birth values in cooled configurations, in dimension one with birth $b\geq 2.0$, in dimension two with birth $b\geq 2.8$. We explicitly checked that the displayed results are approximately independent in shape from the choice of threshold values of comparable sizes. For $\beta\gtrsim 2.6$ the number of dimension two homology classes is compatible with a power-law with exponent $\simeq -22/3$. Linearly mapping inverse couplings squared to temperatures, the latter describes a dilute gas of instantons with semi-classical instanton appearance probability 
\begin{equation}
\exp(-S) = \exp\left(-\frac{8\pi^2}{g^2(T)}\right)\sim \left(\frac{\Lambda_{\mathrm{UV}}}{T}\right)^{b}\,,
\end{equation}
where $\Lambda_{\mathrm{UV}}$ is a UV-cutoff scale and $b=11N_c/3$ from the one-loop beta function for SU($N_c$) gauge theory \cite{larsen2015interacting}, $N_c=2$ in our case. The behavior of the semi-classical instanton-dyon exponentiated action shows the same scaling behavior with $T$ \cite{larsen2015interacting, larsen2016classical}. One-dimensional homology classes with large birth show similar behavior, though with larger deviations from the $T^{-22/3}$ behavior. Likely, they are more affected by thermal fluctuations.

In \Cref{AppendixAngleDifferenceFiltration} we display and discuss birth and persistence distributions of the angle-difference filtration.

\section{Electric and magnetic fields in persistent homology}\label{SecElectrMagnPersHom}

We search for signatures of electric and magnetic screening effects as well as self-duality in superlevel set filtrations of the three gauge-invariant, rotationally invariant quadratic forms constructible from $\EE(x)$ and $\BB(x)$: $\Tr\EE^2(x)$, $\Tr\BB^2(x)$, and the topological density ${q(x)\sim \Tr(\EE(x)\cdot \BB(x))}$.

\begin{figure}
    \centering
	\includegraphics[scale = 0.69]{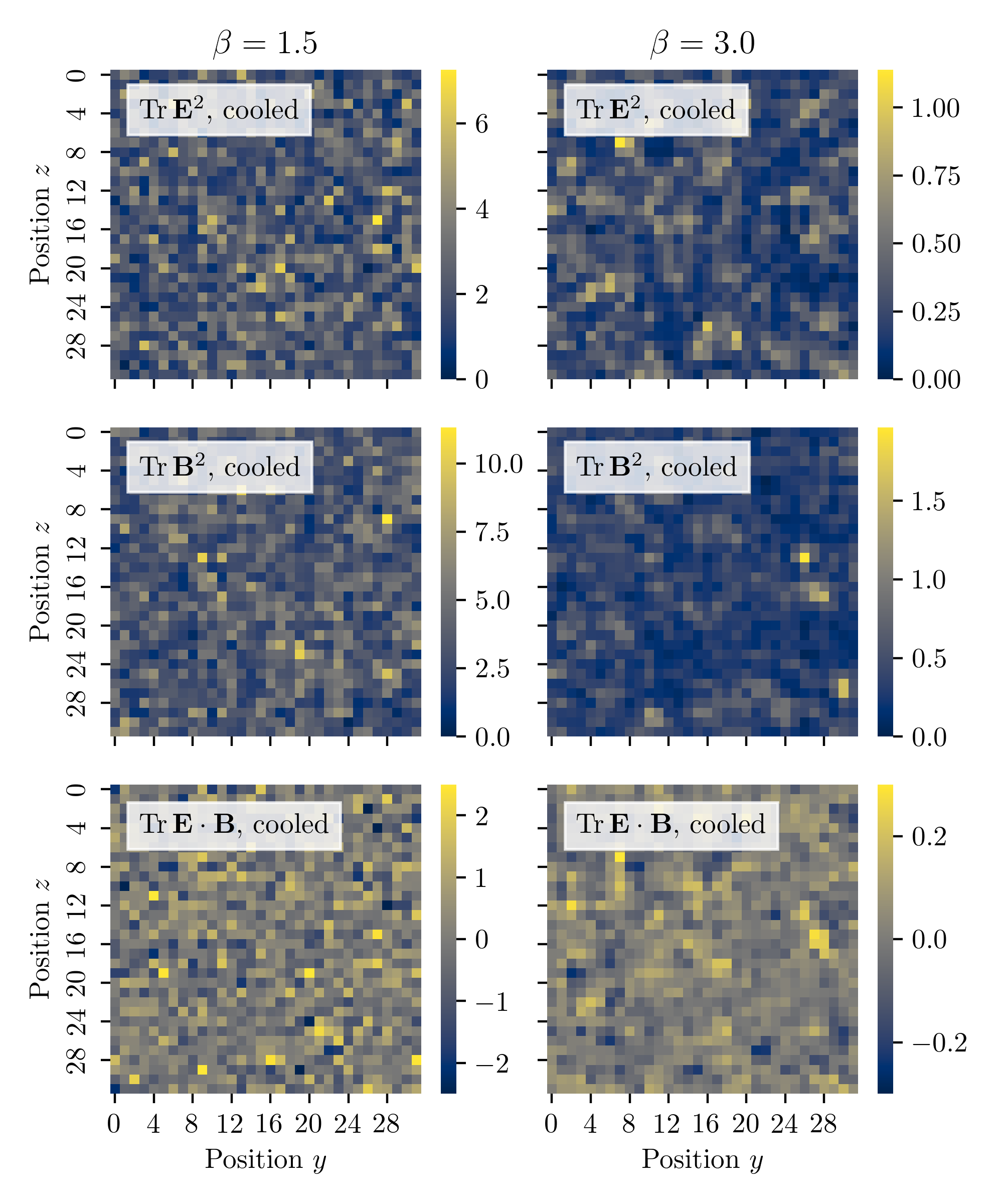}
	\caption{Slices of $\Tr\EE^2(x)$, $\Tr\BB^2(x)$ and $q(x)$ for constant $\xx^1$ and $\tau$ for a single cooled configuration. Data is given in lattice units.}\label{Fig2dslicesEsqrBsqrTopDens}
\end{figure}

On the lattice, SU(2)-valued electric and magnetic fields, denoted $\EE(x)$ and $\BB(x)$, can be defined via antisymmetric averaging of four neighboring plaquettes. For these clover-leaf variants the topological density reads
\begin{equation}\label{EqTopDensity}
q(x) = -\frac{1}{16\pi^2} \Tr(\EE(x)\cdot \BB(x))
\end{equation}
and has well-defined parity, see \Cref{AppendixEBCloverLeaf}. For an impression of $\Tr\EE^2(x)$, $\Tr\BB^2(x)$, and ${\Tr(\EE(x)\cdot \BB(x)) = -16\pi^2 q(x)}$, we display two-dimensional slices in \Cref{Fig2dslicesEsqrBsqrTopDens} for $\beta=1.5$ in the confined and $\beta=3.0$ in the deconfined phase for cooled configurations. Throughout observables, fewer structures are present for $\beta=3.0$ compared to $\beta=1.5$ variants, also smaller by values. This effect is pronounced for $\Tr\BB^2$ compared to $\Tr\EE^2$ and $\Tr\EE\cdot\BB$. Barely any structural changes with $\beta$ are visible without cooling (not shown).

So far, persistent homology quantifiers have been constructed via cubical complexes in three dimensions. Constructions work the same in higher dimensions. For instance, the sublevel set filtration $\Cc_{\Tr\EE^2}$ is constructed by assigning to the 4-cube corresponding to each ${x\in \Lambda}$ the value $\Tr\EE^2(x)$, inductively expanding to lower-dimensional cubes via the lower-star filtration, \Cref{EqLowerStarFiltrationConstruction}. We define the corresponding superlevel set filtration as 
\begin{equation}
\Dd_{\Tr\EE^2}(\nu) := \Cc_{-\Tr\EE^2}(-\nu)\,.
\end{equation}

$\mathrm{Tr}(\mathbf{E}^2(x))$ and $\mathrm{Tr}(\mathbf{B}^2(x))$ are restricted to positive values. For better comparability of low-dimensional persistent homology with unbounded filtrations such as $q(x)$ we investigate their superlevel set filtrations in this section. For unbounded filtrations sub- and superlevel set filtrations are expected to yield similar results. Subsequently, we study topological density superlevel sets.

\subsection{Superlevel sets of $\Tr\EE^2$ and $\Tr\BB^2$}

\begin{figure}
    \centering
	\includegraphics[scale = 0.69]{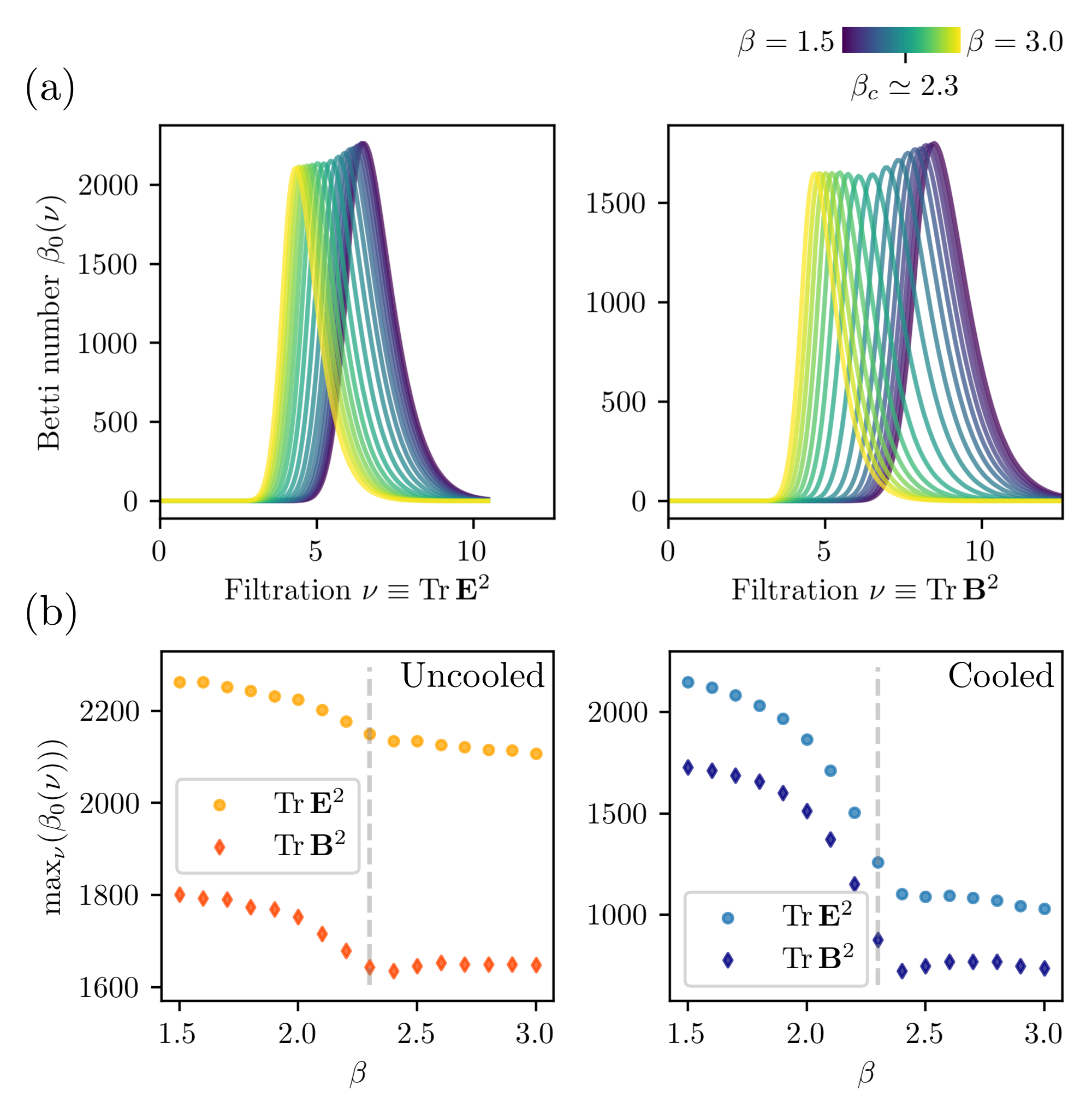}
	\caption{(a): Dimension zero Betti number distributions of $\Tr\EE^2(x)$ (left) and $\Tr\BB^2(x)$ (right) superlevel set filtrations. No cooling has been applied. Data is given in lattice units. (b): Maxima of dimension zero Betti number distributions for uncooled (left) and cooled (right) configurations.}\label{FigEsqrBsqrZerothBettiNumbers}
\end{figure}

We display Betti number distributions of dimension zero homology classes for the superlevel set filtrations of $\Tr\EE^2$ and $\Tr\BB^2$ in \Cref{FigEsqrBsqrZerothBettiNumbers}(a). Higher-dimensional Betti number distributions look similar, see \Cref{AppendixEsqrBsqrBettiCooled}. For cooled Betti number distributions of $\Tr\EE^2$ and $\Tr\BB^2$, we also refer to \Cref{AppendixEsqrBsqrBettiCooled}. For every $\beta$ we observe a distinct peak in both filtrations with positions shifting to lower $\nu$ for increasing $\beta$. This is due to $\langle \Tr\EE^2(x)\rangle$ and $\langle \Tr\BB^2(x)\rangle$ decreasing in the calculations. Starting slowly, shifts are enhanced at intermediate $\beta$, slowing down again for $\beta \gtrsim 2.7$. This is particularly visible in $\Tr\BB^2$. Dimension zero homology classes---in the superlevel set filtration made up by local maxima---occur in the $\Tr\BB^2$ filtration mostly at larger filtration parameters compared to $\Tr\EE^2$. For both $\Tr\EE^2$ and $\Tr\BB^2$, peaks are broadened for low $\beta$.

In \Cref{FigEsqrBsqrZerothBettiNumbers}(b) we compare $\beta$-dependencies of maximal dimension zero Betti numbers of $\Tr\EE^2$ and $\Tr\BB^2$ filtrations, i.e., the maximal number of connected components occurring as the filtration parameter is swept through. Both uncooled and cooled configurations give rise to kink-like behavior at $\beta_c$, indicative of the confinement phase transition. Throughout, we observe a concave decline below $\beta_c$. For $\Tr\EE^2$ the decline continues above $\beta_c$ with a smaller slope; cooling enhances these effects without qualitative changes. For $\Tr\BB^2$ above $\beta_c$ we observe less of a further decline compared to $\Tr\EE^2$, both uncooled and cooled.

Inferred from correlations of $\Tr\EE^2$ and $\Tr\BB^2$ as detailed in \Cref{AppendixEsqrBsqrCorrels}, masses of $\Tr\EE^2$ excitations are larger than for $\Tr\BB^2$ due to electric Debye screening outpacing magnetic screening. Similarly, the mass of electric excitations is expected to be larger than the magnetic mass, resulting in $\langle \Tr\EE^2\rangle < \langle \Tr\BB^2\rangle$ and explaining why magnetic field homology classes persist to larger filtration parameters compared to electric ones. $\Tr\EE^2$ Betti number maxima in \Cref{FigEsqrBsqrZerothBettiNumbers}(b) laying above $\Tr\BB^2$ maxima indicates that $\Tr\EE^2$ sublevel sets contain finer structures, irrespectively of on average higher $\Tr\BB^2$ values. This can already be seen for a single sample in \Cref{Fig2dslicesEsqrBsqrTopDens}. The slow approach of maximal $\Tr\EE^2$ and $\Tr\BB^2$ Betti numbers for growing $\beta \gtrsim \beta_c$ suggests that screening effects discriminate less between electric and magnetic fields at larger inverse couplings. This is again consistent with the approaching masses of $\Tr\EE^2$ and $\Tr\BB^2$ excitations.

Signatures of a higher dominance of self-dual excitations in cooled configurations would manifest in maxima of $\Tr\EE^2$ and $\Tr\BB^2$ which are closer to each other. This is at least not significantly the case and barely visible in the single-sample observables shown in \Cref{Fig2dslicesEsqrBsqrTopDens}.

\begin{figure*}
    \centering
	\includegraphics[scale = 0.69]{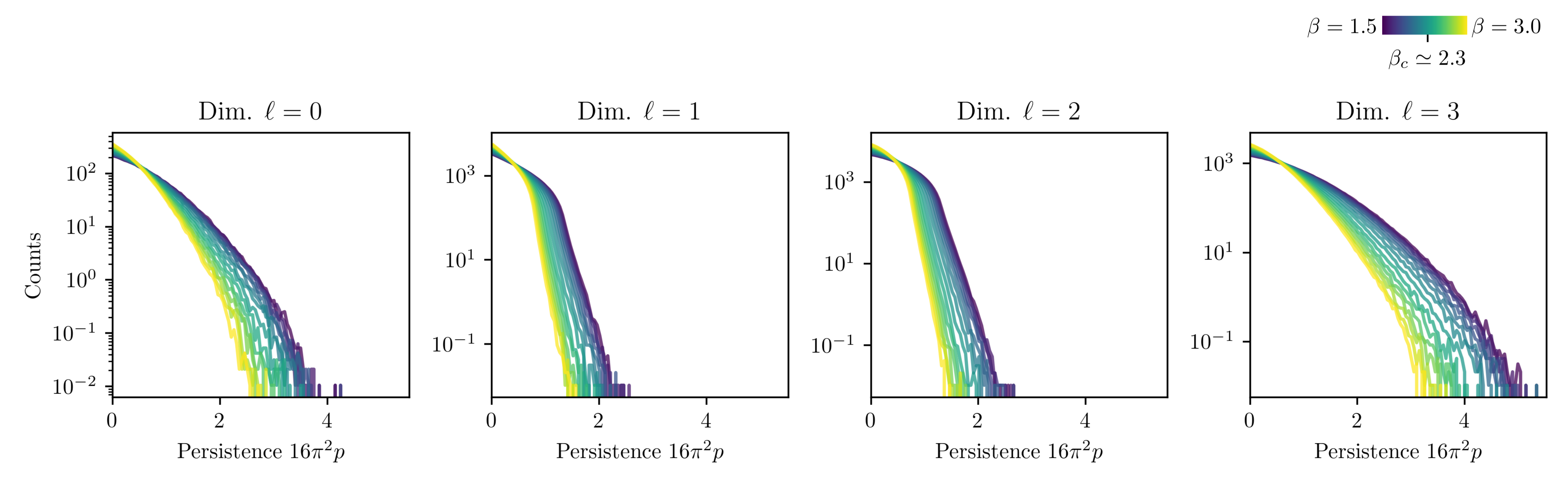}
	\caption{Persistence distributions of the topological density superlevel set filtration from dimensions zero to three. No cooling has been applied; cooled configurations look similar. Results are given in lattice units.}\label{FigTopDensPersistence}
\end{figure*}

\subsection{Superlevel sets of topological densities}\label{SecEBTopDens}
In \Cref{SecPolyakTopDens}, we study the persistent homology of topological densities computed from Polyakov loops, $q_\Pp$. In this section, we discuss the topological density $q(x)\sim\Tr\EE(x)\cdot\BB(x)$, with domain the full four-dimensional lattice $\Lambda$. We show persistence distributions of topological density superlevel sets in \Cref{FigTopDensPersistence} for uncooled configurations. The distributions monotonously decrease for all dimensions and inverse couplings $\beta$. However, their support is reduced to $16\pi^2 p\lesssim 2.5$ for dimensions one and two, while ranging for dimension zero up to $\approx 4.0$ and for dimension three up to $\approx 5.0$. The extended persistence of dimension zero and dimension three homology classes holds for all $\beta$ and indicates that the topological density forms peaks well-separated in space-time, similar to $q_\Pp$ as discussed in \Cref{SecPolyakTopDens}. Cooling barely affects persistence distributions (not displayed). We conclude that lumps of topological density are primarily due to (near-)classical field configurations.

It has been a key finding of \Cref{SecPolyakTopDens} that top-dimensional persistence distributions of $q_\Pp$ follow an exponential distribution which is compatible with instanton-dyon predictions. Such exponential behavior is not visible for $q\sim\Tr\EE\cdot\BB$ as displayed in \Cref{FigTopDensPersistence}. This hints at $q_\Pp$ providing a more suitable lattice approximation to continuous space-time topological densities than $q\sim \Tr\EE\cdot\BB$, less sensitive to temporal fluctuations affected by lattice artifacts. Based on the exact equality of $\int q_\Pp$ and $\int q$ in the continuum, as briefly discussed in \Cref{AppendixTopDensFromPolyak}, we expect that the information content of both topological density variants becomes more similar for larger and finer lattices.

\section{Conclusions}\label{SecConclusions}

We show that persistent homology can be used for unraveling topological structures in the confinement-deconfinement phase transition of finite-temperature Yang-Mills theory with gauge group SU(2). With persistent homology analyses relying on a hierarchical, combinatorial description of input data, a multitude of different cubical complex filtrations constructed from sampled lattice field configurations allows for the extraction of the multifaceted picture of confinement. All filtrations constructed in the present work are gauge-invariant and without an a priori bias towards particular classical configurations. While information about the confinement-deconfinement phase transition may be extracted from a persistent homology analysis of various observables such as the topological density, the traces of electric and magnetic field strengths, and the traced Polyakov loop, we find that amongst these observables the traced Polyakov loop and in particular the related algebra field \labelcref{eq:AlgPol} is best-suited for an application of persistent homology.

The respective Betti numbers potentially represent a novel type of order parameter for this transition, showing kinks at the critical inverse coupling. Strikingly, the analysis reveals that lumps of the topological density formulated in terms of the Polyakov loop dominate the configurations, with persistence statistics indicative of the instanton-dyon occurrence probability. Persistent homology proves crucial for this observation. Filtering by the Polyakov loop algebra field gives access to topological scaling as present in an instanton gas approximation above $T_c$: we show that homology follows the semi-classical scaling arising from the exponential of the instanton action with a (one-loop) running coupling; see \Cref{FigAngleDiffNoHomClasses}. Unlike \cite{Sale:2022qfn}, we do not specifically aim for particular field configurations such as center vortices.

The main results are barely affected by the cooling procedure. Smoothing in general dampens quantum fluctuations beyond a cutoff scale and one has to make sure that the physics of interest does not depend on the cutoff when taking the continuum limit. Our observations suggest that the relevant features are stable against the removal of high-frequency modes, with cooling merely enhancing the associated signatures. Nevertheless, a careful continuum extrapolation and study of the cutoff dependence seems appropriate to confirm that the limit can be taken consistently. Ideally, extracting the features of interest via topological data analysis should not require any smoothing at all. This aspect deserves further investigation and will be the subject of future work, alongside studying the dependence on the ratio of temporal and spatial lattice extents.

The present approach to the topological structure of phase transitions with persistent homology may also support, or benefit from, the application of machine learning techniques. Both supervised and unsupervised learning approaches have been explored for analyzing phase structure \cite{wang2016discovering, van2017learning, carrasquilla2017machine, venderley2018machine, giannetti2019machine, rodriguez2019identifying, carrasquilla2020machine, bachtis2020mapping}. Representing quantum states in variational approaches via neural networks can also be possible \cite{carleo2017solving, carleo2019machine}. Although studies have addressed the issue of extracting explicit expressions for learned order parameters \cite{wetzel2017machine, hu2017discovering, blucher2020towards, wetzel2020discovering}, interpreting neural networks remains challenging. Explainable machine learning techniques, as employed in \cite{blucher2020towards}, are ideally suited for a combined application with persistent homology. Furthermore, gauge-equivariant neural network architectures \cite{kanwar2020equivariant, boyda2021sampling, favoni2022lattice, abbott2022gauge} could make use of the high sensitivity of persistent homology to non-local structures by means of adding appropriate topological layers.

In summary, we demonstrate that the study of homological excitations with persistent homology has great potential for unraveling both dynamical and topological information in QCD and beyond. Particularly appealing is the potential access to topological information without the necessity of cooling as the latter unavoidably removes physical information in the process. A natural step beyond the present work is its extension to the confinement-deconfinement phase transition in SU(3) as well as QCD with dynamical quarks. In short, persistent homology may provide interpretable and accessible order parameters sensitive to strongly suppressed structures in field configurations, yet barely investigated and potentially of high relevance to various areas of research in physics.

\begin{acknowledgments}
We thank J.~Berges, K.~Boguslavski, L.~de~Bruin, V.~Noel, and A.~Wienhard for discussions and work on related projects. This work is funded by the Deutsche Forschungsgemeinschaft (DFG, German Research Foundation) under Germany’s Excellence Strategy EXC 2181/1 - 390900948 (the Heidelberg STRUCTURES Excellence Cluster) and the Collaborative Research Centre, Project-ID No. 273811115, SFB 1225 ISOQUANT. JMU is supported in part by Simons Foundation grant 994314 (Simons Collaboration on Confinement and QCD Strings) and the U.S.\ Department of Energy, Office of Science, Office of Nuclear Physics, under grant Contract Number DE-SC0011090. This work is funded by the U.S.\ National Science Foundation under Cooperative Agreement PHY-2019786 (The NSF AI Institute for Artificial Intelligence and Fundamental Interactions, \url{http://iaifi.org/}).
\end{acknowledgments}

\appendix

\section{Details on the lattice setup}
In this Appendix, we first discuss the HMC setup for our lattice gauge theory calculations. Subsequently, we summarize important aspects of the Wilson flow used for cooling, and provide further details about clover-leaf fields.

\subsection{HMC details}\label{AppendixHybridMonteCarlo}

We employ the standard HMC algorithm to sample field configurations, using 10 leapfrog steps per trajectory with a step size of 0.2. This results in acceptance rates between 60\% and 80\%. First, for each of the considered values of $\beta$, a single Markov chain is initialized with a hot start and then thermalized. Sufficient equilibration is confirmed by observing convergence of the average plaquette and Polyakov loop. 100 samples separated by 10 HMC steps are recorded for each $\beta$ and then further decorrelated in individual Markov chains with 1000 steps each in order to ensure statistical independence of the data.

\subsection{Wilson flow details}\label{AppendixWilsonFlow}

A variety of heuristic algorithms for the smoothing of ultraviolet fluctuations in lattice gauge field configurations has been proposed in the literature, these are commonly called cooling algorithms. The Wilson or gradient flow was introduced as a more rigorous theoretical ansatz to achieve the same goal. The central idea is similar in all approaches, namely minimizing the Wilson gauge action locally in a series of small steps.

For the Wilson flow in particular, this is achieved by numerically solving the gradient flow equation \cite{Bonati:2014tqa}
\begin{align}
    \partial_t U_\mu(x,t) = -g^2 (\partial_{x,\mu}S[U(t)]) U_\mu(x,t)
\end{align}
with a finite step size approximation. Here, $g$ is the bare gauge coupling, $t$ denotes the flow time, and the initial condition $U_\mu(x,0)$ is given by a field configuration obtained through sampling from $\exp(-S[U])$ as described above. The link derivatives are defined in the usual way,
\begin{align}
    \partial_{x,\mu} f(U) &= i \sum_a T^a \frac{\mathrm{d}}{\mathrm{d}s} f(e^{i s X^a} U) \Big|_{s=0} \nonumber \\
    &\equiv i \sum_a T^a \partial_{x,\mu}^{(a)} f(U)\,,
\end{align}
where the $T^a$ are the Hermitian generators of the associated $\mathfrak{su}(N_c)$ algebra (i.e., the Pauli spin matrices for $N_c = 2$), and
\begin{align}
    X^a(y,\nu) =
    \begin{cases}
        T^a &\text{if } (y,\nu) = (x,\mu), \\
        0 &\text{else}.
    \end{cases}
\end{align}
This is equivalent to the computation of the forces in HMC. Throughout this work, the cooled configurations used to compute various results are obtained after 200 flow steps, using a comparably small step size of $\delta t = 0.001$ to avoid discretization errors.

\subsection{Clover-leaf electric and magnetic fields}\label{AppendixEBCloverLeaf}
Clover-leaf electric fields are the SU(2) elements
\begin{widetext}
\begin{align}
E_i(x):=&\; \frac{1}{4} \mathrm{Im}\bigg[U_{4i}(x) + U^\dagger_{i}(x-\hat{i})U_{4i} (x-\hat{i})U_{i}(x-\hat{i}) + U^\dagger_{4}(x - \hat{4})U_{4i}(x-\hat{4})U_{4}(x-\hat{4})\nonumber\\
&\qquad\quad  + U^\dagger_{4}(x-\hat{4})U^\dagger_{i}(x-\hat{i}-\hat{4})U_{4i}(x-\hat{i}-\hat{4})U_{i}(x-\hat{i}-\hat{4})U_{4}(x-\hat{4})\bigg]\,,\label{EqEavDef}
\end{align}
which transform under a local gauge transformation $V(x)$ as $E_i(x)\mapsto V(x)E_i(x)V^\dagger(x)$. Clover-leaf magnetic fields are the SU(2) elements
\begin{align}
B_i(x) =&\; \frac{1}{8}\varepsilon^{ijk}\mathrm{Im}\bigg[U_{jk}(x)  + U_j^\dagger(x-\hat{j})U_{jk}(x-\hat{j}) U_j(x-\hat{j}) +  U_k^\dagger(x-\hat{k})U_{jk}(x-\hat{k})U_k(x-\hat{k}) \nonumber\\
&\qquad\qquad\;  +U_k^\dagger(x-\hat{k})U_j^\dagger(x-\hat{j}-\hat{k})U_{jk}(x-\hat{j}-\hat{k})U_j(x-\hat{j}-\hat{k})U_k(x-\hat{k})\bigg]\,,\label{EqBavDef}
\end{align}
\end{widetext}
transforming as $B_i(x)\mapsto V(x)B_i(x)V^\dagger(x)$. 

By spatial antisymmetrization the clover-leaf topological density,
\begin{equation}\label{EqDefEuclideanTopDensity}
q(x) = -\frac{1}{2^9 \pi^2}\sum_{\mu\nu\rho\sigma=\pm 1}^{\pm 4} \tilde{\varepsilon}^{\mu\nu\rho\sigma} \Tr(U_{\mu\nu}(x)U_{\rho\sigma}(x))\,,
\end{equation}
has a well-defined parity \cite{di1981preliminary,rothe2012lattice}. The fully antisymmetric $\tilde{\varepsilon}^{\mu\nu\rho\sigma}=\tilde{\varepsilon}_{\mu\nu\rho\sigma}$ is defined through $1 = \tilde{\varepsilon}_{1234} = -\tilde{\varepsilon}_{2134} = -\tilde{\varepsilon}_{(-1)234}$. Plaquettes for negative directions equate to
\begin{subequations}\label{EqPlaquettesNegativeDirections}
\begin{align}
U_{(-\mu)\nu}(x) =&\;U^\dagger_{\mu}(x-\hat{\mu})U^\dagger_{\mu\nu}(x-\hat{\mu})U_\mu(x-\hat{\mu}),\label{EqPlaqMinusMu}\\
U_{\mu(-\nu)}(x) =&\;U^\dagger_{\nu}(x-\hat{\nu})U^\dagger_{\mu\nu}(x-\hat{\nu})U_\nu(x-\hat{\nu}),\label{EqPlaqMinusNu}\\
U_{(-\mu)(-\nu)}(x)=&\; U^\dagger_\nu(x-\hat{\nu})U^\dagger_{\mu}(x-\hat{\mu}-\hat{\nu})\nonumber\\
\times U_{\mu\nu}(x&-\hat{\mu}-\hat{\nu})U_\mu(x-\hat{\mu}-\hat{\nu})U_\nu(x-\hat{\nu})\,.\label{EqPlaqMinusMuNu}
\end{align}
\end{subequations}
A straight-forward lattice computation confirms \Cref{EqTopDensity} for $q(x)$ in terms of $E_i(x)$ and $B_i(x)$.

\section{The mathematics of persistent homology}\label{SecAppendixMathPersDiags}
In this Appendix we discuss details of the construction of cubical complexes as well as their homology and persistent homology groups. Often in persistent homology so-called simplicial complexes are used, which are in a way triangular variants of cubical complexes.

\subsection{Cubical complexes}
Based on \cite{wagner2012efficient}, cubical complexes are mathematically defined as follows. 
A cubical complex is constructed from elementary closed intervals of non-degenerate, $[n,n+1]$, or degenerate type, $[n,n]$, $n\in \nn$. Chains of intervals are formal superpositions of them. The boundaries of intervals are defined as the chains ${\partial[n,n+1] = [n+1,n+1]-[n,n]}$ and $\partial[n,n]=0$. An elementary cube $C$ is a Cartesian product of such elementary intervals, $C=I_1\times\dots \times I_m$, where each $I_n$ may be non-degenerate or not. The dimension of $C$ is defined to be the number of non-degenerate intervals present in the product. We refer to $0$-cubes as vertices, $1$-cubes as edges and $2$-cubes as squares. The boundary $\partial C$ of a cube $C$ is the chain
\begin{align}
\partial C =&\; (\partial I_1\times I_2\times\dots\times I_d) + (I_1\times\partial I_2\times \dots \times I_d) \nonumber\\
&\; + \dots + (I_1\times I_2 \times\dots\times \partial I_d)\,.
\end{align}
Given these constructions, a cubical complex $\Cc$ is a collection of elementary cubes closed under taking boundaries this way. A cubical complex $\Cc$ has dimension $d$ if $d$ is the maximal dimension of the cubes $C\in\Cc$.

\subsection{Homology groups}
A cubical complex $\Cc$ being a topological space, its homology groups $H_\ell(\Cc)$ may be investigated. Their dimensions $\dim H_\ell(\Cc)$ specify the number of $\ell$-dimensional homological features present in the complex $\Cc$. We consider homology groups with coefficients in $\zz_2$. Corresponding thus directly to the intuitive notion of chains of $\ell$-cubes, we define chain groups $C_\ell(\Cc)$ as the free $\zz_2$-modules over the $\ell$-chains present in $\Cc$. The already specified boundary operator $\partial$ extends to chains of cubes, $\partial_\ell: C_\ell(\Cc)\to C_{\ell-1}(\Cc)$. We find $\partial_\ell\circ \partial_{\ell+1}=0$, such that the chain groups naturally form a chain complex; we may define the cycle group $Z_\ell(\Cc)=\mathrm{ker}\; \partial_\ell$ and the boundary group $B_\ell(\Cc) = \mathrm{im}\; \partial_{\ell+1}$. For any cubical complex we find $B_\ell(\Cc)\subseteq Z_\ell(\Cc)$ as subgroups, such that their quotient may be taken. Homology groups are then defined as
\begin{equation}
H_\ell(\Cc):= Z_\ell(\Cc)/B_\ell(\Cc)\,.
\end{equation}
Technically, the $\ell$-th homology group contains equivalence classes of $\ell$-cycles (closed $\ell$-chains of cubes) modulo boundaries of $(\ell+1)$-cycles. This directly allows for the association of homology classes as corresponding to independent holes of dimension $\ell$ present in the cubical complex of interest. The $\zz_2$-dimension of $H_\ell(\Cc)$ counts their number and is called the $\ell$-th Betti number,
\begin{equation}
\beta_\ell(\Cc) := \dim_{\zz_2} (H_\ell(\Cc))\,.
\end{equation}

\subsection{Persistent homology groups}
Given a filtration of cubical complexes, i.e., a family $(\Cc_r)_{r\in\rr}$ of cubical complexes with $\Cc_r \subseteq \Cc_s$ for $r\leq s$, their individual homology groups $(H_\ell(\Cc_r))_r$ may be computed. In addition, the inclusion maps $\Cc_r\to \Cc_s$, $r\leq s$, induce maps on homology group level 
\begin{equation}
\iota^{r,s}_\ell:H_\ell(\Cc_r)\to H_\ell(\Cc_s)\,, 
\end{equation}
whenever $r\leq s$. Such a map maps an $\ell$-dimensional homology class present in $\Cc_r$ either to an $\ell$-dimensional homology class in $\Cc_s$ or to zero, indicating that the homology class is not present anymore in $\Cc_s$. Also, the $\iota_\ell^{r,s}$ can have non-trivial cokernels: new homology classes can appear in $\Cc_s$, not in the image of previous maps. Then, for any $\epsilon > 0$:
\begin{equation}\label{EqNewHomolClass}
H_\ell(\Cc_{s-\epsilon})\subsetneqq H_\ell(\Cc_s)\,.
\end{equation}
The collection $(H_\ell(\Cc_r),\iota^{r,s}_\ell)_{r\leq s,\ell}$ is called a persistence module. If \Cref{EqNewHomolClass} is true only for finitely many distinct $s$ we call the persistence module tame.

The structure theorem of persistent homology, proven in \cite{edelsbrunner2000topological, zomorodian2005computing}, states that any tame persistence module is isomorphic to a persistence diagram, i.e., a finite multiset of birth-death pairs $(b,d)\in \rr^2$ with $b < d$. In a multiset the same elements may appear multiple times. A birth-death pair $(b,d)$ corresponds to an independent hole present in the complexes $\Cc_r$ for $r\in [b,d)$.

Different metrics are available on the space of persistence diagrams, most importantly the so-called Bottleneck, Wasserstein and interleaving distances. With respect to all of these stability theorems have been established \cite{cohen2007stability, cohen2010lipschitz, bauer2013induced}, implying that if input data changes slightly, then persistence diagrams computed from this data are also perturbed only slightly.

Although averages of persistence diagrams as multisets cannot be defined unambiguously \cite{mileyko2011probability}, generic other persistent homology quantifiers have this property \cite{bubenik2015statistical}. If persistent homology observables are evaluated on a lattice of finite extent such as in our work, their volume-averages converge towards well-defined large-volume limits \cite{hiraoka2018limit, spitz2020self}.

\section{Correlation functions}
In this Appendix we discuss two-point correlation functions of traced Polyakov loops $P(\xx)$ and electric and magnetic fields squared. This will provide insights into emergent screening masses and infrared behavior related to confinement.

We focus on connected correlation functions and their $y$-, $z$- and $\tau$-direction zero-modes. We define
\begin{align}
    \bar{E}^2(n_x):=&\;\frac{1}{N_\sigma^2 N_\tau}\sum_{n_y,n_z,n_\tau}\big[\Tr\EE^2(n_x,n_y,n_z,n_\tau)\nonumber\\
    &\qquad\qquad\qquad\qquad  - \langle \Tr\EE^2(n_x,n_y,n_z,n_\tau)\rangle\big]\,,
\end{align}
maintaining fluctuations depending on $n_x$. Then, correlations are computed as
\begin{equation}
    \langle |\Tr\EE^2(p_x)|^2\rangle_c = \sum_{x} \bar{E}^2(n_x)\bar{E}^2(0)e^{-i p_x n_x}
\end{equation}
for $n_x\in \{0,\dots,N_\sigma-1\}$ and
\begin{equation}
p_x \in \left\{-\pi, -\frac{(N_\sigma-2)\pi}{N_\sigma},\dots,\frac{(N_\sigma-2)\pi}{N_\sigma}\right\}\,,
\end{equation}
and analogously for $\Tr\BB^2(x)$ and $q(x)$. For $P(\xx)$ correlations, temporal averaging is trivial. Correlations are displayed depending on physical momenta on the lattice,
\begin{equation}
\tilde{p}_x = 2 \sin\left(\frac{p_x}{2}\right)\,.
\end{equation}
Due to the connectedness of the shown correlators, values at $p_x=0$ are largely suppressed. Zero modes of the correlators including disconnected contributions would give rise to susceptibilies.

\subsection{Polyakov loop correlations}\label{AppendixPolyakovCorrels}
\begin{figure}
    \centering
	\includegraphics[scale = 0.69]{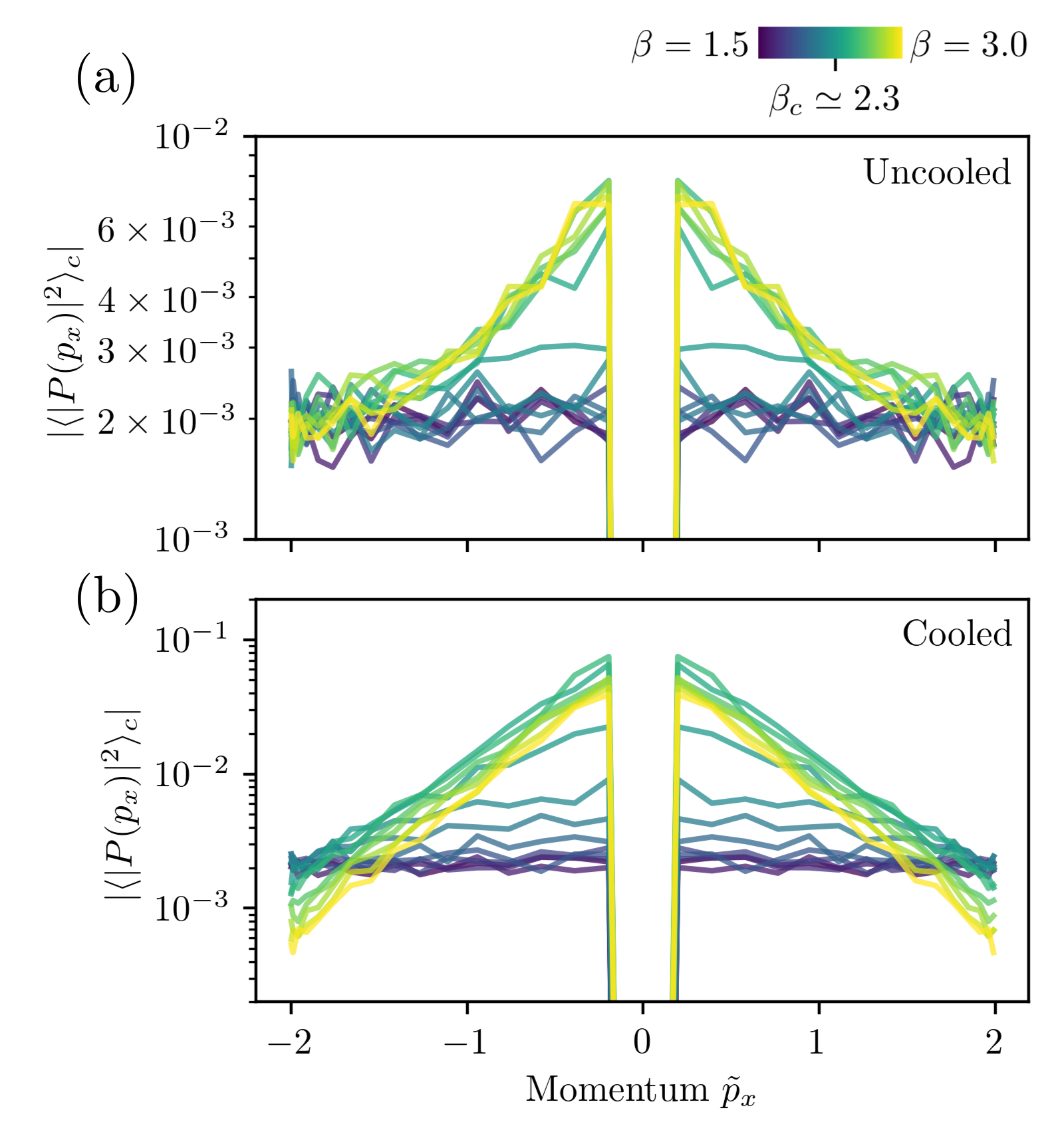}
	\caption{Connected Polyakov loop two-point correlation function $|\langle |P(p_x)|^2\rangle_c|$ for (a) uncooled and (b) cooled configurations. Data is given in lattice units.}\label{FigPolyakovConnCorrelator}
\end{figure}

In \Cref{FigPolyakovConnCorrelator} we display Polyakov loop correlations for uncooled (a) and cooled configurations (b) for the entire $\beta$-range. A qualitative change can be observed around $\beta_c$ for uncooled configurations. Below $\beta_c$, the correlator is constant in $\beta$ and in $\tilde{p}_x$ up to fluctuations, indicative of free quasi-particles dominating $P(\xx)$ \cite{diakonov2007confining}. Above $\beta_c$, a peak emerges around the $\tilde{p}_x=0$ mode, indicative of $P(\xx)$ excitations acquiring a screening mass due to spontaneous center symmetry breaking. Above an intermediate regime up to $\beta\simeq 2.5$, correlations stay approximately constant. They level off for large momenta to baseline fluctuations.

Below $\beta_c$, cooling smoothens the baseline fluctuations, leaving their height unaltered. Thus, they are due to (near-)classical configurations. With cooling, for smaller $\beta$ than before a peak emerges in the deep infrared, until it fully developed for $\beta\simeq 2.6$. For such couplings, the momentum-dependence of the peak-tails is approximately exponential. The peak decreases in overall height for large $\beta$ above 2.7, potentially indicative of the reduction of instantons with increasing temperatures as observed already in \Cref{SecAngleDiffFiltration}.

\begin{figure}
    \centering
	\includegraphics[scale = 0.69]{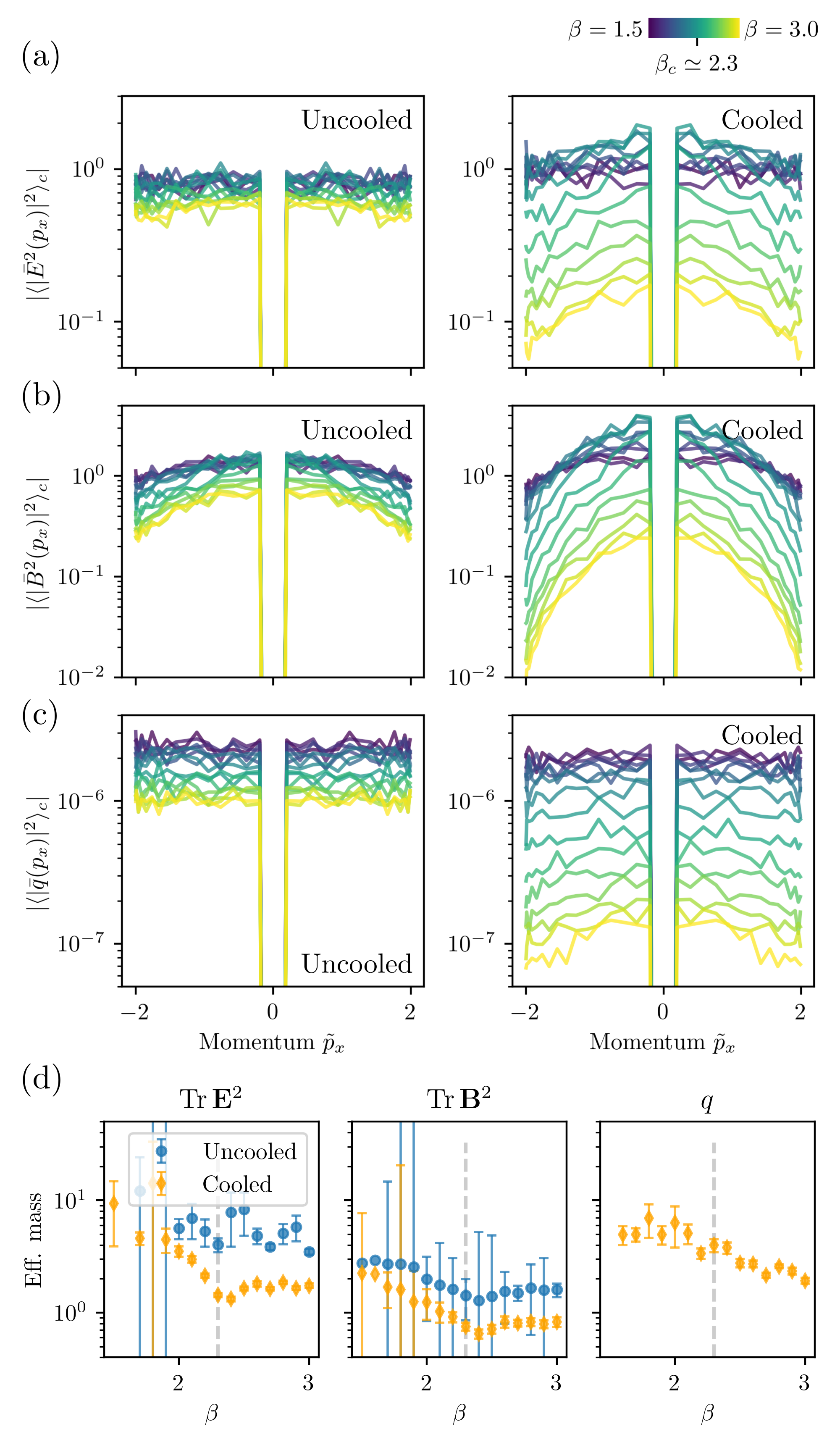}
	\caption{Connected two-point correlations of (a) electric and (b) magnetic field strengths, as well as (c) topological densities for uncooled (left) and cooled (right) configurations. The dip at $\tilde{p}_x=0$ is due to (approximately homogeneous) disconnected contributions removed. (d): Effective masses of $\Tr\EE^2$ (left), $\Tr\BB^2$ (center) and $q$ (right) excitations, deduced from a least-squares fit to a Lorentz curve $\sim m/(\tilde{p}_x^2+m^2)$. Zero modes are excluded from fits; errorbars are extracted from the least-squares fit covariance matrix. Fits did not converge for uncooled $q$. Data is given in lattice units.}\label{FigEsqrBsqrCorr}
\end{figure}

\subsection{Correlations of electric and magnetic excitations}\label{AppendixEsqrBsqrCorrels}
In \Cref{FigEsqrBsqrCorr}(a) to (c) we display connected two-point correlation functions of electric and magnetic fields squared, $\Tr\EE^2$ and $\Tr\BB^2$, as well as of the topological density $q\sim \Tr\EE\cdot\BB$. First discussing configurations without cooling, barely any $\beta$-dependence is visible at low $\beta$ near 1.5. For intermediate values of $\beta$ correlations decrease in value, slowing down again for $\beta$-values near 3.0. Electric fields give rise to a slightly enhanced infrared peak at larger inverse couplings squared. For magnetic fields this effect is visible more clearly. Topological density correlations are nearly flat for the entire $\beta$-range.

The consequences of cooling are similar to the Polyakov loop correlator. Deviations from the constant behavior for $\beta$ near 1.5 start to occur for $\beta\gtrsim 1.8$. Any behavior present for larger inverse couplings is pronounced by cooling. This is due to more dominant (near-)classical field configurations after cooling.

All this can be understood from effective masses of the excitations. We extract masses from correlators using a least-squares fit to a Lorentz curve $\sim m/(\tilde{p}_x^2+m^2)$. In \Cref{FigEsqrBsqrCorr}(d) we display correspondingly fitted masses, comparing cooled and uncooled ones. Cooling reduces masses, since less ultraviolet fluctuations enter self-energies and effectively generate the masses. Masses of $\Tr\EE^2$ excitations are observed to be consistently larger than masses of $\Tr\BB^2$ excitations --- a direct consequence of Debye screening. In addition, a kink is visible in effective masses near $\beta_c$, suggesting that the relevant structures for confinement enter self-energies, too. Fits of $q$-masses did not converge for uncooled configurations. Masses of $q$ excitations after cooling lay somewhat between those of $\Tr\EE^2$ and $\Tr\BB^2$ excitations. Their comparable height can be regarded as a signature of self-dual configurations playing a role.

\section{Polyakov loop topological densities: miscellanea}\label{AppendixTopDensFromPolyak}
In this Appendix, we first discuss the rewriting of winding numbers using Polyakov loops, giving rise to the Polyakov loop topological density $q_\Pp$. Subsequently, we describe cooled birth and persistence distributions of the $q_\Pp$ filtration.

\subsection{Rewriting winding numbers with Polyakov loops}
In this Appendix we briefly discuss the rewriting of the winding number
\begin{equation}
Q_{\mathrm{top}} = \frac{1}{32\pi^2}\int_{T^4}\varepsilon_{\alpha\beta\mu\nu}\Tr F_{\alpha\beta} F_{\mu\nu}
\end{equation}
in terms of the Polyakov loop $\Pp:T^3\to \mathrm{SU}(2)$, based on \cite{ford1998monopoles}. We take the 4-torus $T^4$ to have extents $N_x,N_y,N_z,N_\tau$, analogously to the lattice of interest. 

Starting from gauge potentials $A_\mu$ on the 4-torus $T^4$, transition functions defined on the entire $\rr^4$ are defined via the periodicity properties of $T^4$, which manifests for all $x\in T^4$ and $\mu=1,\dots,4$ in
\begin{equation}
A_\mu(x+N_\nu) = U^{-1}_\nu(x) A_\mu(x) U_\nu(x) + iU_\nu^{-1}(x)\partial_\mu U_\nu(x)\,.
\end{equation}
The transition functions fulfil the cocycle condition 
\begin{equation}
U_\mu(x)U_\nu(x+N_\mu) = U_\nu(x)U_\mu(x+N_\nu)
\end{equation}
and transform under a gauge transformation $V(x)$ as
\begin{equation}
U_\mu^V(x) = V^{-1}(x) U_\mu(x) V(x+N_\mu)\,.
\end{equation}

Suppose the transition functions satisfy ${U_i(\xx,\tau=0)=1}$ for all $i=1,2,3$ and $U_4 = 1$. Then, skipping derivation steps detailed in \cite{ford1998monopoles}, we find
\begin{equation}
Q_{\mathrm{top}}=\frac{1}{24\pi^2}\int_{B_4} \varepsilon_{0ijk}\Tr[(\Pp^{-1}\partial_i \Pp)(\Pp^{-1}\partial_j \Pp)(\Pp^{-1}\partial_k \Pp)]\,,
\end{equation}
where $B_4=\{(\xx,\tau)\in T^4\,|\, \tau=0\}$.

\subsection{Cooled birth and persistence distributions}
\begin{figure*}[p]
    \centering
	\includegraphics[scale = 0.69]{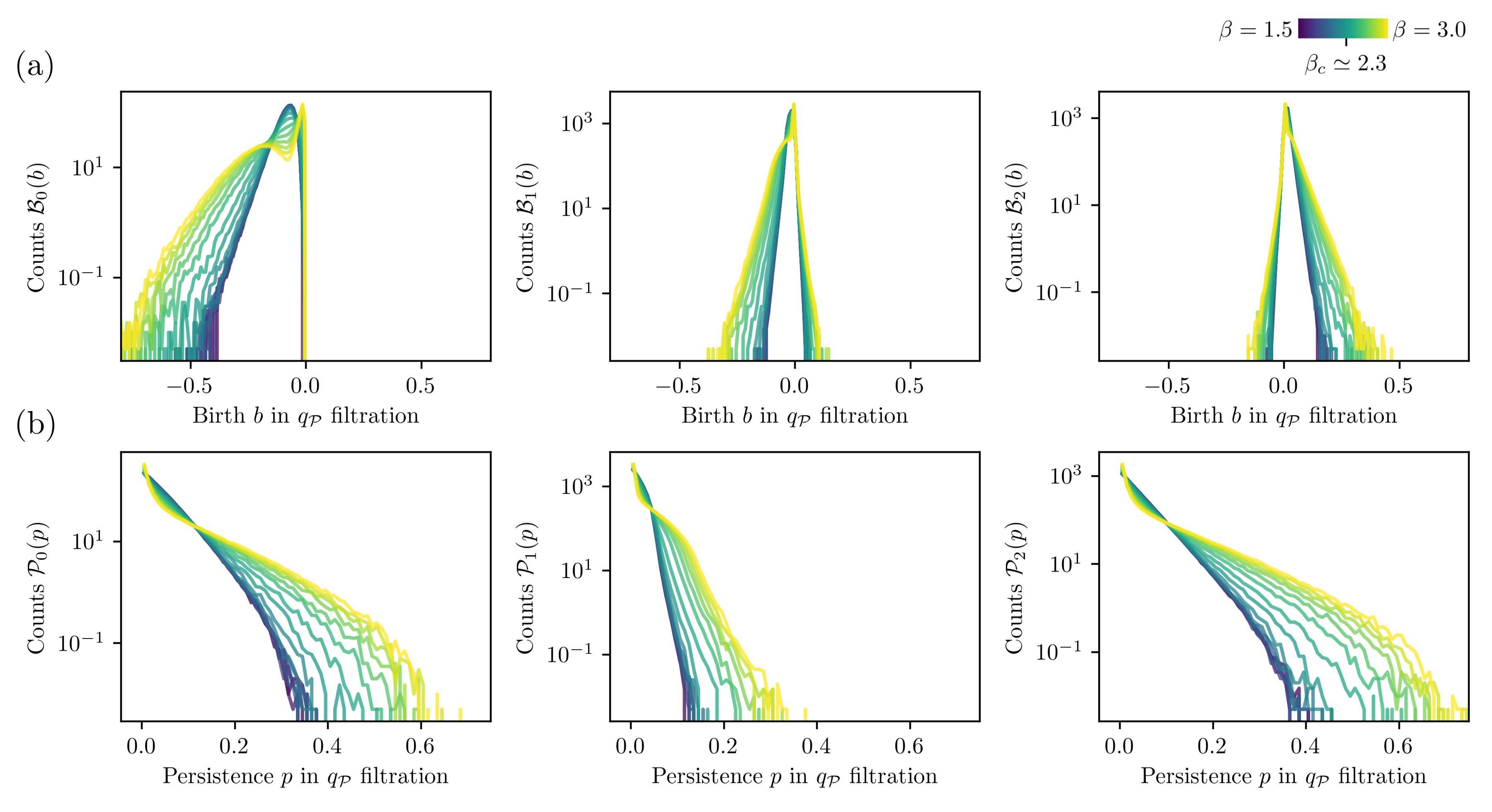}
	\caption{Homological quantifiers of Polyakov loop topological density $q_\Pp$ sublevel sets. (a): Birth distributions for dimensions zero to two. (b): Persistence distributions for dimensions zero to two. Data is given in lattice units for cooled configurations.}\label{FigAppendixPolyakTopDensBirthPersCooled}
\end{figure*}

In \Cref{FigAppendixPolyakTopDensBirthPersCooled} we display birth and persistence distributions of the Polyakov loop topological density sublevel set filtration for cooled configurations. Comparing to \Cref{FigPolyakovTopDens}, where uncooled variants are shown, we note that below $\beta_c$ cooled distributions are similar to uncooled ones though persistence distributions have larger support. Above $\beta_c$, major deviations occur. The broadening of uncooled birth distributions transforms after cooling into an additional peak in dimension zero and a novel shoulder in dimension one after cooling. Dimension two birth distributions above $\beta_c$ reveal larger broadening towards positive $q_\Pp$-values. This goes along with persistence distributions after cooling spreading more towards larger persistences compared to uncooled data. 

While the similarity of cooled and uncooled $q_\Pp$-structures below $\beta_c$ indicates that topological densities are dominated by (near-)classical configurations, above $\beta_c$ ultraviolet fluctuations show up more often. Cooling reveals additional topological structures above $\beta_c$.

\section{Angle-difference filtration: Birth and persistence distributions}\label{AppendixAngleDifferenceFiltration}
\begin{figure*}[p]
    \centering
	\includegraphics[scale = 0.69]{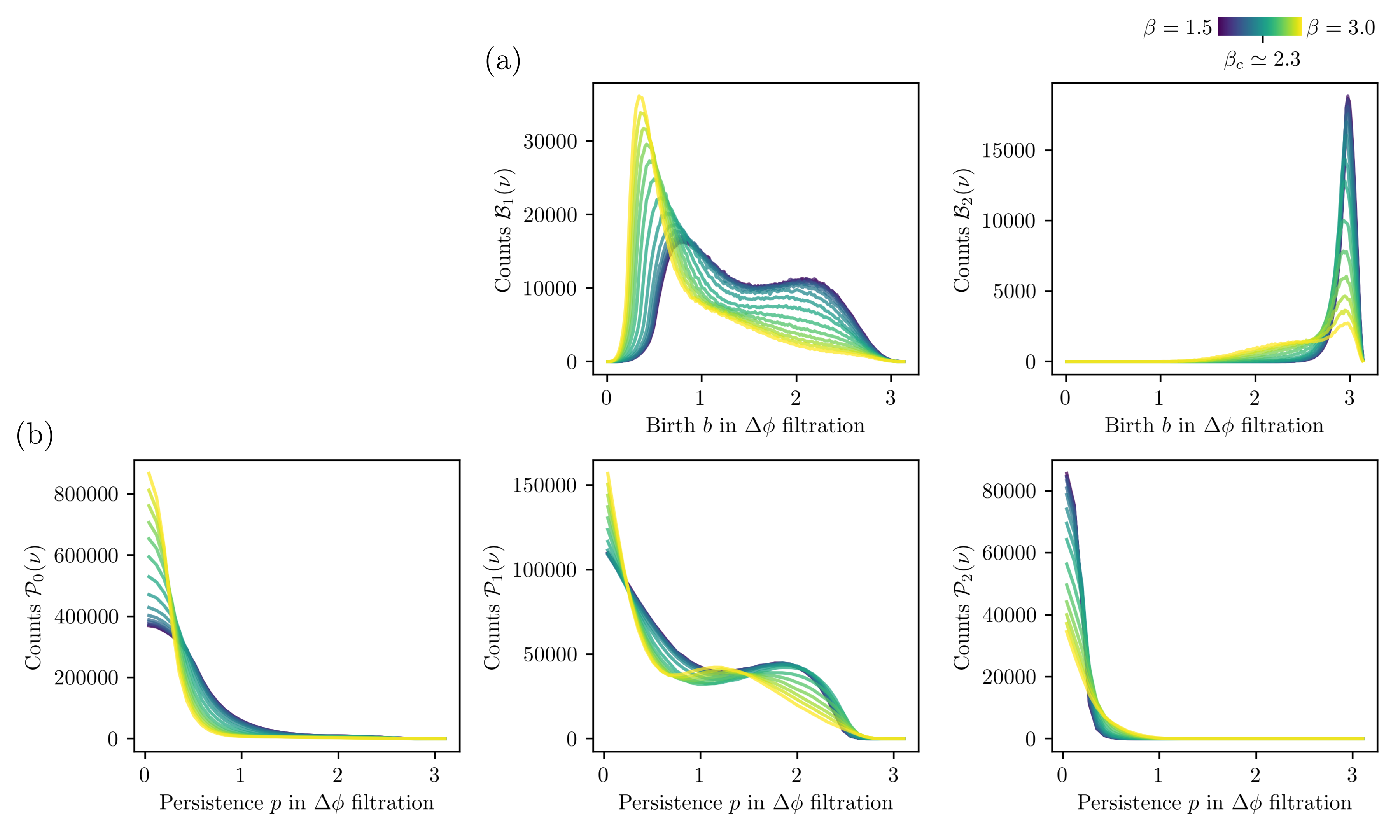}
	\caption{Homological quantifiers of the angle-difference filtration. (a): Birth distributions for dimensions one and two. All connected components are born at zero in the angle-difference filtration, thus dimension zero birth distributions are not displayed. (b): Persistence distributions for dimensions zero to two. Data is shown for cooled configurations.}\label{FigAppendixAngleDiffBirthPers}
\end{figure*}

In \Cref{FigAppendixAngleDiffBirthPers}(a) we display birth  distributions of the angle-difference filtration of the holonomy Lie algebra field $\phi(\xx) = \arccos ( P(\xx))$ for cooled configurations. Zero-dimensional birth distributions are not displayed, since they are by construction trivial: all dimension zero homology classes are born at filtration parameter zero. Dimension one birth distributions at $\beta\approx 1.5$ have two peaks: one near $b\approx 0.9$ and a second near $b\approx 2.1$. The latter strongly diminishes above $\beta_c$, while the former gets enhanced. The lower-$\Delta\phi$ peak emerges in dimension two homology classes only above $\beta_c$. Below, a single large peak near $\Delta\phi\approx 3.0$ is present, which strongly decreases in height for $\beta \gtrsim \beta_c$.

Persistence distributions are shown in \Cref{FigAppendixAngleDiffBirthPers}(b) for cooled configurations. Persistences of dimension zero homology classes in the angle-difference filtration monotonously decrease, though with smaller $\beta$-dependence for lower $\beta\approx 1.5$.  In dimension one we see a second peak near $\Delta\phi \approx 2.0$ at low $\beta$, which vanishes above $\beta_c$ and gives rise to a persistence peak near $\Delta\phi\approx 1.2$. Noise results again in a large number of homology classes with very low persistences. Persistences of dimension two homology classes are mostly very low, which follows from comparably large birth parameters and the phase difference bound $\Delta\phi \leq \pi$.

\section{Persistent homology of $\Tr\EE^2$, $\Tr\BB^2$ and $q$ superlevel sets}\label{AppendixEsqrBsqrBettiCooled}
In this Appendix we first discuss the persistent homology for all dimensions zero to three of gauge-invariant electric and magnetic field quadratic forms. A comparison with cooled configurations follows.

\subsection{Betti numbers of uncooled configurations}
We display dimension zero to dimension three Betti number distributions of the superlevel set filtrations of $\Tr\EE^2$ and $\Tr\BB^2$ for uncooled configurations in \Cref{FigAppendixEsqrBsqrTopDensUncorr}(a) and (b). The topological density Betti number distributions shown in \Cref{FigAppendixEsqrBsqrTopDensUncorr}(c) are discussed below. For every $\beta$ we observe a single peak, shifting to lower filtration parameters with increasing dimension. This is due to the superlevel set filtration. Decreasing the filtration parameter $\nu$, at first maxima appear as dimension zero homology classes. A multitude of these with saddle points in between is required to form dimension one homology classes; they get born at lower $\nu$. This trend continues to higher dimensions. Finally, dimension three homology classes (enclosed 3-volumes) die if $\nu$ reaches corresponding minimum values.

For both $\Tr\EE^2$ and $\Tr\BB^2$ peak positions shift to lower $\nu$ with increasing $\beta$ due to $\langle \Tr\EE^2(x)\rangle$ and $\langle \Tr\BB^2(x)\rangle$ decreasing in simulations. Topological structures of any dimension occur in the $\Tr\BB^2$ filtration mostly at larger filtration parameters compared to $\Tr\EE^2$. Across dimensions and for both $\Tr\EE^2$ and $\Tr\BB^2$, peaks are broadened for low $\beta$. All this is qualitatively similar to the dimension zero Betti numbers discussed in \Cref{SecElectrMagnPersHom} and consistent with electric and magnetic screening masses as deduced in \Cref{AppendixEsqrBsqrCorrels}.

Topological densities not bounded from below as electric and magnetic fields squared, their Betti number distributions in \Cref{FigAppendixEsqrBsqrTopDensUncorr}(c) are not limited to positive filtration parameters. Instead, dimension two distributions have support around zero filtration parameters, dimension three distributions mostly at negative filtrations parameters. This is indicative of local topological density values scattering symmetrically around zero. Maximal values reveal kink-like behavior around $\beta\approx \beta_c$ and in overall numbers are comparable to the $\Tr\EE^2$ filtration.

\subsection{Betti numbers of cooled configurations}
\begin{figure*}[p]
    \centering
	\includegraphics[scale = 0.69]{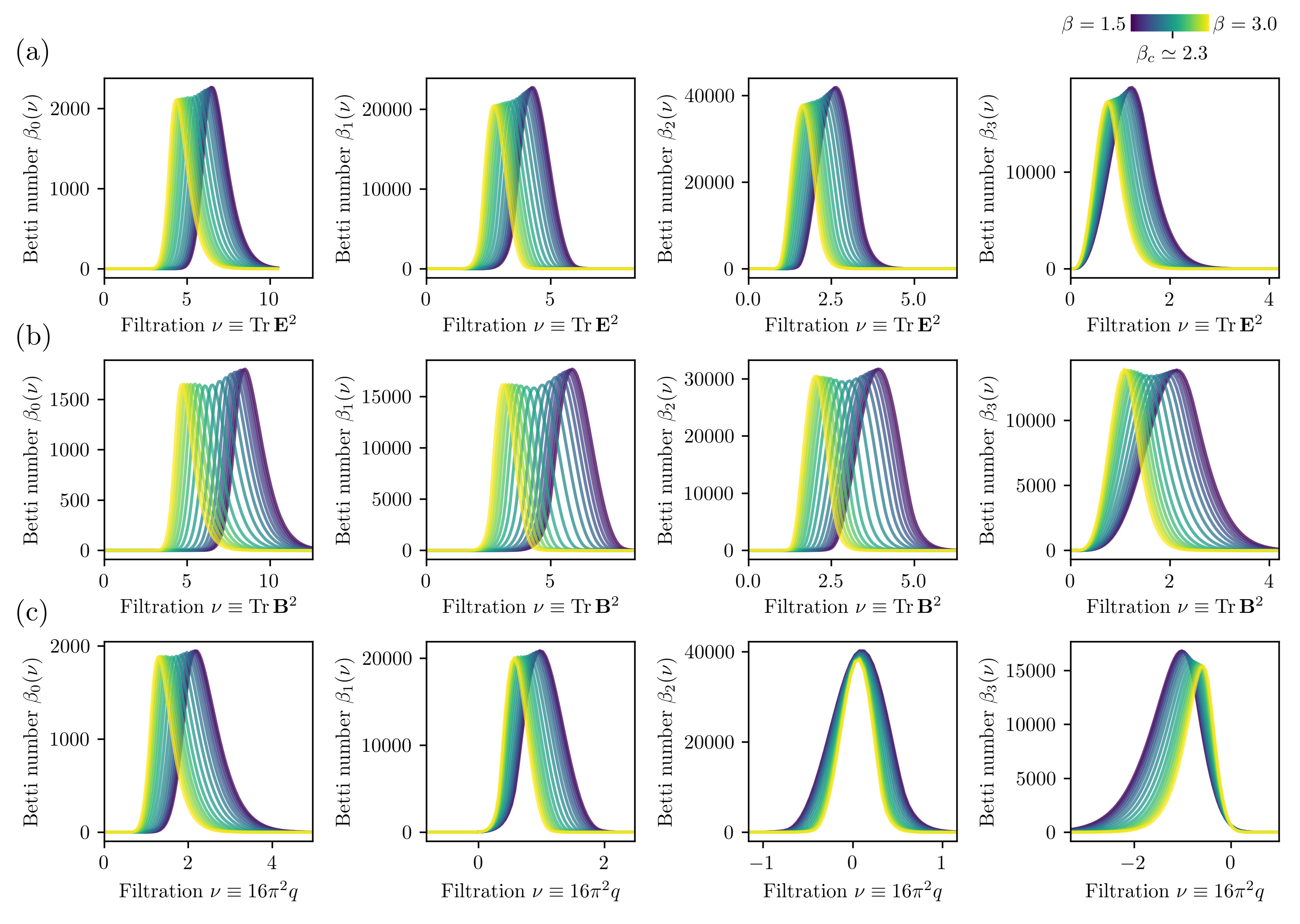}
	\caption{Betti number distributions in dimensions zero to three of (a) $\Tr\EE^2$, (b) $\Tr\BB^2$, (c) $q$ superlevel set filtrations for uncooled configurations. Data is given in lattice units.}\label{FigAppendixEsqrBsqrTopDensUncorr}
\end{figure*}

\begin{figure*}[p]
    \centering
	\includegraphics[scale = 0.69]{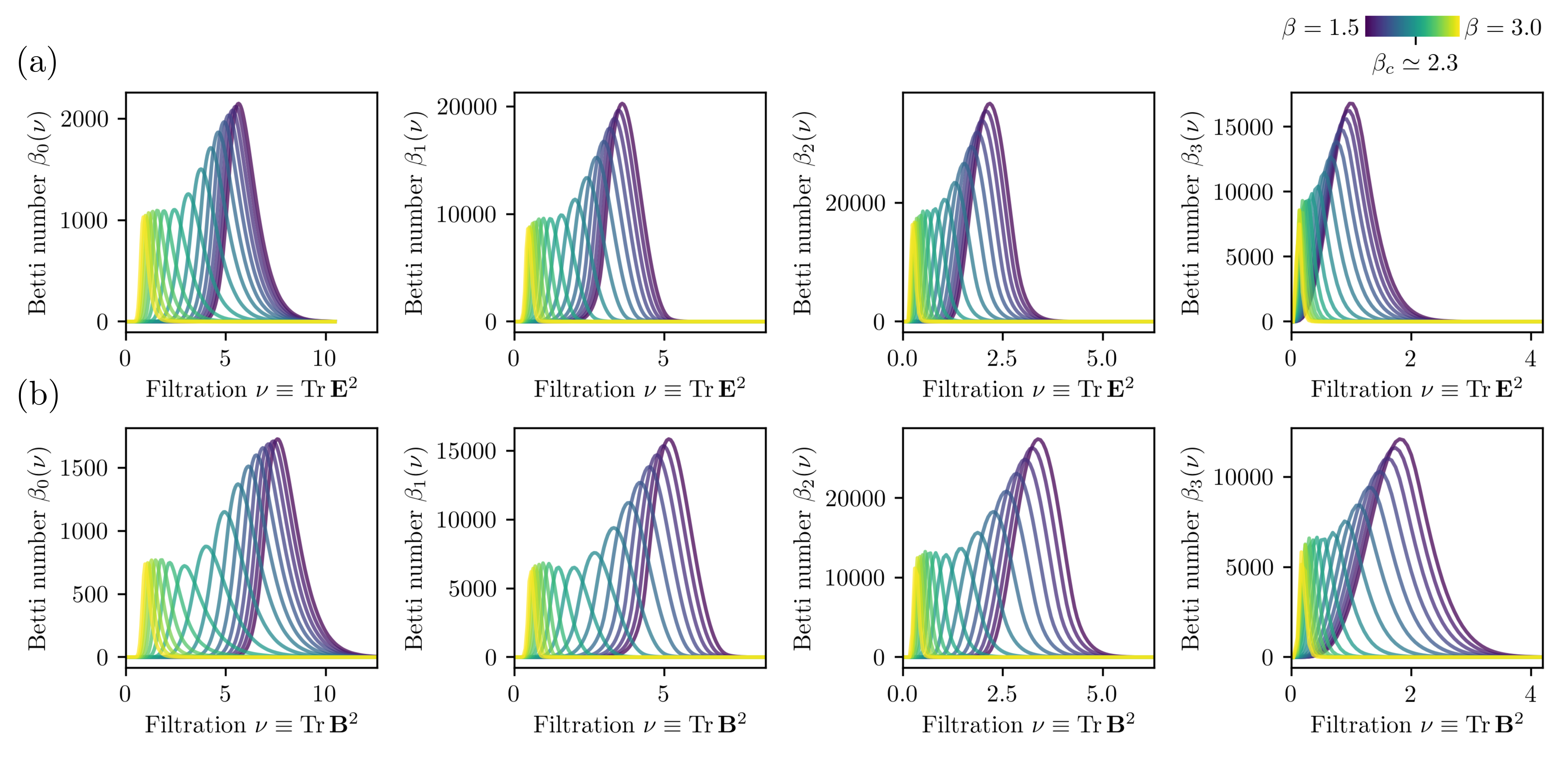}
	\caption{Betti number distributions in dimensions zero to three of (a) $\Tr\EE^2$ and (b) $\Tr\BB^2$ superlevel set filtrations for cooled configurations. Data is given in lattice units.}\label{FigAppendixEsqrBsqrCooled}
\end{figure*}

For cooled configurations we show Betti number distributions of all dimensions for $\Tr\EE^2$ and $\Tr\BB^2$ superlevel set filtrations in \Cref{FigAppendixEsqrBsqrCooled}. Similar plots have been shown in \Cref{FigAppendixEsqrBsqrTopDensUncorr}(a) and (b) for configurations without cooling. Upon comparison, we see that cooling has barely any effect for low $\beta\approx 1.5$. However, after cooling and for larger $\beta$ the number of homology classes is reduced compared to uncooled configurations. Qualitative changes across all dimensions occur near $\beta_c$. For $\beta\gtrsim \beta_c$ maxima of Betti number distributions saturate in height as indicated already in \Cref{FigEsqrBsqrZerothBettiNumbers}(b). Cooled structures move to very small filtration parameters compared to uncooled ones.

\bibliography{literature}

\end{document}